\documentclass{icrc29}
\usepackage{graphicx,amssymb,amsmath,times}
\begin{document}
\title[The Status of VHE Gamma-Ray Astronomy]
{The Status of VHE Gamma-Ray Astronomy}
\author[Rene A. Ong] {Rene A. Ong\\
        Department of Physics
	and Astronomy, 
	University of California, Los Angeles \\
	Los Angeles,
	CA 90095, USA \\
	}
\presenter{Presenter: R.A. Ong (rene@astro.ucla.edu).}

\maketitle

\begin{abstract}

\quad\qquad I present a summary of the status of the field of very
high-energy (VHE) gamma-ray astronomy, as of
early 2006.
This paper is based on the
Rapporteur Talk given at the 29th International 
Cosmic Ray Conference (ICRC 2005)
in Pune, India (August 2005).
It covers astrophysical results from observations made
by high-energy and very high-energy telescopes operating
at photon energies above 1 GeV.
The majority of recent observations in this field have been made
by ground-based telescopes using the atmospheric Cherenkov
or air shower techniques.

In this paper, I try to accomplish two tasks:
first, to present a broad overview of the many papers presented
at ICRC 2005 in the sessions on gamma-ray astronomy, and
second, to provide a status report of the field that
highlights the key results that have been reported 
in the last two years (since ICRC 2003 \cite{Mori}).
Some subjectivity is unavoidable in the selection of
results to present in summaries of this type, and
there are interesting results that are not discussed here.

A number of results presented at ICRC 2005 were preliminary
in nature and 
have been subsqeuently updated with a journal submission.
In addition, there have been a number of significant new results
that have been reported since the conference.
Wherever possible, I have attempted to reference 
journal submissions or publications that have occurred
since ICRC 2005.

\end{abstract}

\section{Introduction}
This paper
summarizes the presentations made at ICRC 2005 in
the following sessions:

\begin{enumerate}  
\item OG 2.1:  Diffuse X-rays, $\gamma$-rays.
\item OG 2.2:  Galactic sources (SNRs, pulsars, etc.).
\item OG 2.3:  Extragalactic sources (AGN, clusters, etc.).
\item OG 2.4:  Gamma-ray bursts.
\item OG 2.7:  New experiments and instrumentation.
\end{enumerate}     

\noindent A total of 207 papers were presented in these sessions;
they were approximately divided equally between oral and poster
presentations.
At ICRC 2005, there were relatively few contributions
relating to X-ray astronomy and to $\gamma$-ray astronomy below
1\,GeV.
Thus, to a very large extent, the core subject area covered in
these sessions was $\gamma$-ray astronomy above 1 GeV
(and particularly the observational results from ground-based
telescopes operating in the very high energy (VHE) regime above
50 GeV).
This paper concentrates on the results presented from
currently operating telescopes.
Due to space constraints,
significantly less emphasis is placed on
instrumentation, although there is a brief discussion towards
the end of the paper on the major, upcoming future telescopes.

\section{Experimental Summary}
The spectrum of high-energy photons is divided into the X-ray and $\gamma$-ray
bands, which typically overlap at photon energies of 100-200\,keV.
The major X-ray satellite instruments operating at the present time
are RXTE, ASCA, Chandra, and XMM-Newton.
In the $\gamma$-ray band between 100 keV and 100 MeV, the operating
satellite telescopes are HETE-2, Swift, and INTEGRAL.
At this meeting, there was a comprehensive summary of the scientific
results so far obtained with the INTEGRAL mission \cite{Produit}.
There is currently no operating space-based telescope with significant
sensitivity to high-energy $\gamma$-rays above 100 MeV.
The last major telescope in this band was EGRET on the Compton Gamma
Ray Observatory (CGRO) which operated between 1991 and 2000.

Gamma rays at energies above 100 GeV have been typically divided into
the Very High Energy (VHE) and Ultra High Energy (UHE) bands
\cite{Weekes,Ong}.
However, times change, and with so much interest
now focused on the GeV and TeV wavebands, it is simplest to define
a single VHE band for all energies $E > 50\,$GeV.
Currently, all telescopes with significant sensitivity in the VHE
band are ground-based instruments that use the atmospheric Cherenkov
or air shower techniques.

\begin{table}   
\caption{\label{expt-summary} 
Currently operating VHE gamma-ray telescopes.
The name of each telescope is given, along with its
type (AC=Atmospheric Cherenkov), location, altitude, specifications,
and reference at this meeting.  The specifications list the
currently installed detector area (mirror area for atmospheric Cherenkov
and instrumented detector area for air shower).
There is no reference at this meeting for the CACTUS telescope.}
\vspace{0.3cm}
\begin{center}
\begin{tabular}{||c|l|c|l|c|l||} \hline \hline
Experiment & Type & Location & Altitude & Specifications & Ref. \\
\hline
& & & & & \\
CACTUS & AC-Sampling & Barstow, USA
& 640\,m & 144 x 42\,m$^2$ & { } \\
CANGAROO-III & AC-Imaging & Woomera, Australia 
& 165\,m & 4 x 78\,m$^2$ & \cite{Yoshikoshi} \\
HESS & AC-Imaging & Gamsberg, Namibia
& 1800\,m & 4 x 110\,m$^2$ & \cite{Hoffman} \\
MAGIC & AC-Imaging & La Palma, Spain 
& 2250\,m & 1 x 226\,m$^2$  & \cite{Mirzoyan} \\
PACT & AC-Sampling & Pachmarhi, India  
& 1075\,m & 25 x 4.5\,m$^2$ & \cite{Bose} \\
SHALON & AC-Imaging & Tien Shan, Kazakhstan 
& 3338\,m & 1 x 11\,m$^2$ & \cite{Sinitsyna1}\\
STACEE & AC-Sampling & Albuquerque, USA 
& 1700\,m & 64 x 37\,m$^2$ & \cite{Kildea1}\\
TACTIC & AC-Imaging & Mt. Abu, India
& 1400\,m & 1 x 9.5\,m$^2$ & \cite{Rannot}\\
VERITAS & AC-Imaging & Mt. Hopkins, USA 
& 1275\,m & 2 x 110\,m$^2$ & \cite{Holder}\\
Whipple & AC-Imaging & Mt. Hopkins, USA 
& 2250\,m & 1 x 78\,m$^2$ & \cite{Perkins} \\
& & & & & \\
\hline
& & & & & \\
ARGO-YBJ & Air Shower &  Yangbajing, Tibet 
& 4300\,m & 4000\,m$^2$ & \cite{Vernetto}\\
GRAPES-III & Air Shower & Ooty, India
& 2200\,m & 288 x 1\,m$^2$ & \cite{Grapes}\\
Milagro & Air Shower & Los Alamos, USA
& 2630\,m & 4800\,m$^2$ & \cite{Smith1} \\
Tibet & Air Shower & Yangbajing, Tibet 
& 4300\,m & 761 x 0.5\,m$^2$ & \cite{Sakata}\\
& & & & & \\
\hline
\end{tabular}
\end{center}
\end{table}

A gamma ray entering the Earth's atmosphere interacts with an air
molecule to produce an electron-positron pair.
This pair radiates photons via the bremsstrahlung process and
an electromagnetic cascade develops, creating an air shower
in the atmosphere.
Charged particles (mostly electrons) in the air shower radiate
Cherenkov light that is beamed to the ground with a cone opening
angle of $\sim 1.5^\circ$.
{\em Atmospheric Cherenkov Telescopes} detect VHE $\gamma$-rays 
by capturing the rapid ($\sim 5\,$ns) Cherenkov flashes amidst
the background of night sky photons.
These telescopes use large optical mirrors to focus the mostly
blue Cherenkov radiation onto fast photomultiplier tubes (PMTs).
The primary advantages of the atmospheric Cherenkov technique
are high sensitivity, excellent angular resolution and energy
resolution, and relatively low energy threshold.
The disadvantages are moderate duty-cycle ($\sim$ 10\%) and small
field-of-view (FOV).
The Cherenkov telescopes operating today include CACTUS,
CANGAROO-III, HESS, MAGIC, PACT, SHALON, STACEE, TACTIC, VERITAS.
and Whipple.
CACTUS, PACT, and STACEE are examples of wavefront-sampling
telescopes that use an array of mirrors to gather the Cherenkov
radiation, measuring the arrival time and amplitude of the Cherenkov
pulse at many distributed locations on the ground.  The
other telescopes are examples of the more established imaging
Cherenkov technique where the Cherenkov radiation is focused onto
an imaging camera at one or more locations on the ground.
See Table~1 for a summary of the atmospheric Cherenkov telescopes.

Some fraction of the charged particles and photons in VHE air showers
reach the ground level and can be detected by {\em Air Shower Telescopes}.
These telescopes typically consist of charged particle detectors
(scintillators or resistive plate counters) spread out on a grid 
and often covered by a lead
layer to convert the photons in the shower, or water Cherenkov
detectors in which a large tank of water is viewed by a number of
fast PMTs.
The primary advantages of the air shower technique are high
duty-cycle and very wide FOV.
Disadvantages are moderate sensitivity,
energy resolution and angular resolution, and relatively high
energy threshold.
Thus, the two major ground-based techniques for detecting 
VHE $\gamma$-rays are fully complementary -- both techniques
have proven essential in exploring the VHE sky.
The air shower telescopes operating today include
ARGO-YBJ, GRAPES-III, Milagro, and Tibet (see Table 1).

\section{Scientific Highlights}
ICRC 2005 was an exciting and very fruitful conference, with many
new results presented.  Rather than simply enumerate all the many
papers in an encyclopedic fashion, it is valuable to highlight
the most noteworthy scientific results that were presented.

The most compelling new results in the area of VHE astrophysics
presented this year are the following:
\begin{enumerate}  
\item The discovery of many new sources in the Galactic plane.
      HESS reported results from a deep survey of the galactic plane,
      and from subsequent follow-up observations, that present
      evidence for a significant number of new sources.  Some of
      these new sources can be well-correlated with known astronomical
      objects, but other sources cannot be easily identified.
      An important new source whose $\gamma$-ray
      processes are not understood is the
      Galactic Center.
\item Detailed studies made of galactic sources.  
      For the first time,
      a number of sources have been well mapped in terms of their spectra
      and their spatial extent.  The observational work has been carried
      out by atmospheric Cherenkov telescopes, especially HESS,
      but there has been, in addition, considerable progress made on the
      theoretical side.  
\item The discovery of four new active galactic nuclei (AGN) by HESS and
      MAGIC and measurements of AGN properties by a number of
      atmospheric Cherenkov telescopes.  Three of new AGN sources are in
      the redshift range of 0.15-0.20, making them the most distant
      extragalactic sources yet detected at very high energies -- this
      could well have implications for our understanding of the
      extragalactic background light.
\item Other new discoveries and measurements.
      Significant new results are the discovery of diffuse radiation 
      in the galactic plane and of broad emission in 
      the Cygnus region by Milagro.
      The galactic plane result represents the first detection of diffuse 
      emission at very high energies. 
      In addition to these results, there are a number of interesting
      measurements of properties of known sources and searches for
      new galactic and extragalactic sources.
\end{enumerate}     

In the following sections, we present the results from the first
item above - the HESS Galactic plane survey and then discuss the
new results divided up by source class.

\subsection{HESS Galactic Plane Survey}

HESS reported results from their survey of the central
region of the Galactic plane \cite{Funk}.
The survey was carried out in 2004
with the completed four-telescope array, and it covered
a region in Galactic longitude from l = -30$^\circ$ to
l = 30$^\circ$.  The coverage in Galactic latitude was
approximately b = $\pm 3^\circ$.
HESS used $\sim 230\,$hrs to carry out the survey
(including additional time for follow-up observations 
in the Galactic Center, RX J1713-3946, and other regions).
The average flux sensitivity of the survey was 3\% of the
Crab Nebula at energies above 200 GeV.
Eleven sources were detected with a post-trials statistical
significance greater than six standard deviations ($> 6\sigma$).
Of these sources, only two (Galactic Center and RX J1713)
had been reported previously at very high energies.
In addition, HESS reported seven new sources with a post-trials
statistical significance greater than four standard 
deviations \cite{Aharonian}.
Figure~1 shows the significance map for the HESS survey.

\begin{figure}[h]
\begin{center}
\includegraphics*[width=0.80\textwidth,angle=0,clip]{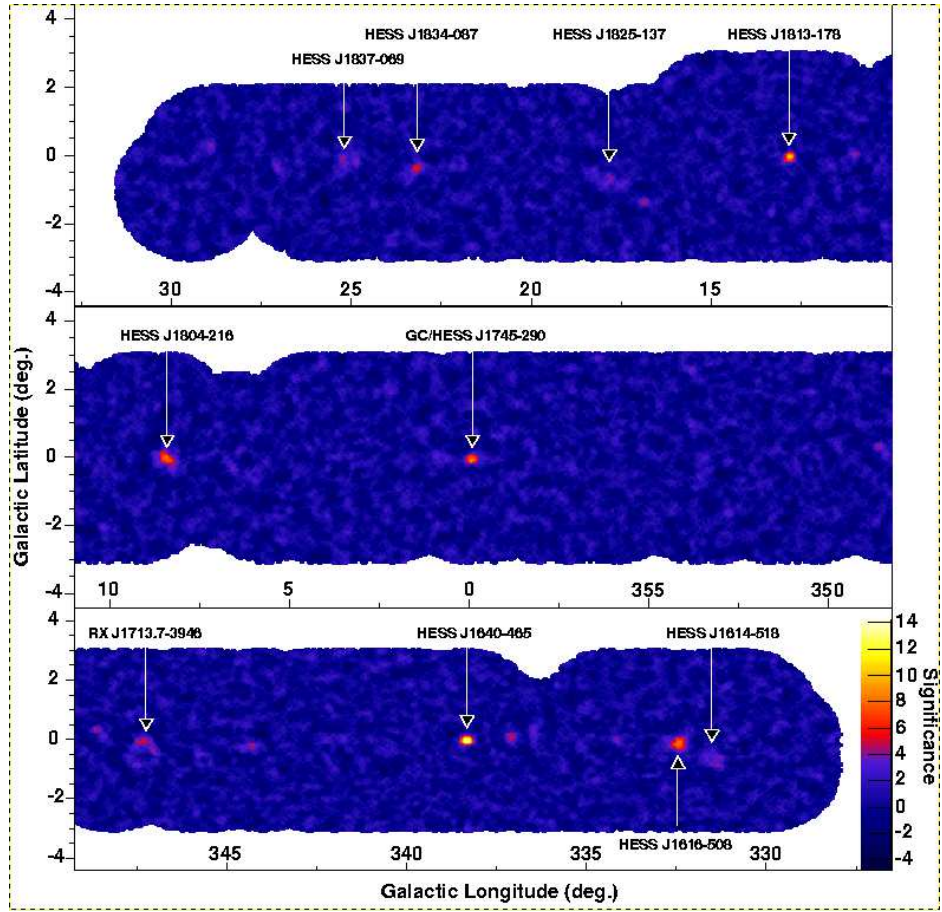}
\caption{\label {fig1} 
  Significance map of the HESS Galactic plane survey carried out
in 2004 \cite{Funk}.
The survey covered 60$^\circ$ in Galactic longitude
(horizontal axis)
and approximately $\pm 3^\circ$ in Galactic latitude (vertical axis).
Eleven sources were detected in the survey with a statistical
significance greater than six standard deviations,
as labeled in the figure.
More complete results on additional data, including more discovered
sources, have been recently reported \cite{Aharonian}.}
\end{center}
\end{figure}

There are several strong indications that the sources detected
by HESS have a galactic, rather than extragalactic, origin.
First, the sources are concentrated along the galactic plane
with a mean Galactic latitude of -0.17$^\circ$.
Second, the width of the Galactic latitude distribution 
is consistent with the
distribution of young pulsars and supernova remnants
in the Galaxy.
Finally, all of the sources are extended beyond the size
of the HESS point spread function, a result that would rule
out, for example, an AGN component.
The supernova remnant RX J1713-3946 is the source observed
with the largest spatial extent.

We expect VHE $\gamma$-rays to be produced as a result of extreme
non-thermal particle acceleration.
In principle, the observed $\gamma$-ray emission can come
from non-thermal bremsstrahlung or inverse-Compton scattering
processes involving relativistic electrons or from the decays
of neutral pions produced from the interactions of protons
and nuclei with ambient material.
Potential sources in our Galaxy include pulsars and 
pulsar wind nebulae (PWN), supernova remnants (SNRs), 
microquasars, and regions associated with massive star formation. 
It is obviously essential to correlate the VHE sources
detected in the Galactic plane with known objects in order to
establish the various source classes and to search for new,
and unexpected, types of VHE $\gamma$-ray emitters.
The HESS group has made a systematic study of possible counterparts
for the sources detected in their survey -- a summary of
this study at the time of ICRC 2005 is shown in Table~2.
(More complete information can be found in \cite{Aharonian}).

\begin{table}   
\caption{\label{hess-association} 
Possible counterparts to the sources detected in the
HESS survey of the Galactic plane.
The first column gives the
sources listed by the HESS source name.
The second column gives the possible counterpart
(SNR=supernova remnant, PWN=pulsar wind nebula,
 and UNID= unidentified source from EGRET, ASCA or INTEGRAL).
The Association column gives a subjective determination
of the strength or likelihood of the association.
The Reference column identifies the corresponding 
paper submitted to this meeting.
All sources were detected with a significance greater than
six standard deviations except
the source HESS J1826-148, marked by an asterisk.
}
\vspace{0.3cm}
\begin{center}
\begin{tabular}{||l|l|l|l||} \hline \hline
HESS Source & Possible Counterpart & Association & Ref. \\
\hline
& & & \\
J1713-397 & SNR RX J1713.7-3946 & Firm & \cite{Berge,Berge_update} \\
J1745-290 & Galactic Center (SGR A) & Firm & \cite{Rolland} \\
J1747-281 & SNR G0.9+0.1        & Firm & \cite{Khelifi}\\
J1826-148$^*$ & $\mu$-Quasar LS 5039 & Firm & 
\cite{deNaurois,deNaurois_update}\\
& & & \\
\hline
& & & \\
J1614-518 & - & Unknown & \cite{Funk} \\
J1616-508 & PWN (PSR J1617-5055) & Tentative & \cite{Funk} \\
J1640-465 & SNR/UNID (G338.3-0.0) & Tentative & \cite{Funk} \\
J1804-216 & SNR (G8.7-0.1)/PWN (PSR J1803-2137) & V. Tentative & \cite{Funk} \\
J1813-178 & SNR (GG12.8-0.0) & Tentative & \cite{Funk} \\
J1825-137 & PWN/UNID (PSR J1826-1334) & Tentative & 
        \cite{deJager,deJager_update} \\
J1834-087 & SNR (G23.3-0.3) & Tentative & \cite{Funk} \\
J1837-069 & UNID (AX J1838.0-0665) & Tentative & \cite{Funk} \\
& & & \\
\hline
\end{tabular}
\end{center}
\end{table}

As shown in Table~1, four of the sources from the survey are
firmly identified with known objects: SNR RX J1713-3946, the Galactic
Center (SGR A), SNR G0.9+0.1, and the microquasar LS5039.
Of the remaining sources, five have possible associations with
supernova remnants or pulsar wind nebulae, one (HESS J1837-069)
has a possible association with an unidentified ASCA X-ray source,
and two sources (HESS J1614-518 and HESS J1804-216)
do not have any well-motivated association.

In addition to using the angular overlap with known objects,
the identification of the HESS sources can be further explored
using the measured VHE spectra and using the correlation with
maps of atomic and molecular constituents.
The latter study is important for understanding if the VHE
emission is correlated with sites of enhanced interstellar matter
density, a possibility that would be expected if, for example,
gamma rays result from the collision of accelerated cosmic
rays with clouds of interstellar material.
The typical spectrum for the sources is hard, with an
average spectral index of $\alpha \sim 2.3 $,
an observation that generally supports the notion of Fermi
acceleration.
An early, but extremely interesting, comparison of the survey
results with maps of CO and HI was presented, indicating
possible association with molecular clouds along the line of
sight in the case of a few sources \cite{Lemiere}.

\section{OG 2.1: Diffuse $\gamma$-ray Sources}

The majority of the $\gamma$-rays detected by EGRET 
can be ascribed to diffuse radiation from the interactions of
cosmic rays with gas and dust in the disk of the Galaxy.
This galactic diffuse $\gamma$-radiation is generally well modeled
at energies below 1 GeV, but above 1 GeV there is a well known
discrepancy between the observational data and the model
predictions, with the data showing a statistically significant flux
excess.  There has been a great deal of speculation in the literature
as to the origin of the ``GeV excess'', covering the range from 
astrophysics (a population of new high-energy sources or harder
spectra for the cosmic-ray protons or electrons) to nuclear
physics (modified interaction cross-sections for the cosmic-ray
collisions with material in the Galactic plane) to particle physics
(dark matter annihilations).
At ICRC 2005 there
were relatively few submissions on this topic,
perhaps because many in the community now realize that the launch
of GLAST in two years should substantially improve the observational
situation; one paper suggests an important contribution to the
excess comes from X-ray binaries \cite{Bhatt} and a second outlines
the expected emission from supernova remnants \cite{Satyendra}.

Although there is not yet a generally agreed upon explanation for
the EGRET GeV excess, the importance of measurements made by
ground-based instruments is widely acknowledged.
Such measurements provide the important extrapolation of the EGRET
results to very high energies and could, in principle, lead to
a map of the diffuse radiation with much higher angular resolution than
EGRET.  Several VHE telescopes reported results from observations regions
of the Galactic plane. Here, we first discuss the results 
from observations of the plane as a whole, and then
we turn our attention to results from observations of specific
regions in the plane.

\subsection{Galactic Plane -- Overall}
Milagro, a water Cherenkov air shower array operating at a median energy
of $\sim 3.5$\,TeV, reported on observations made over a 
three year period of two regions in the Galactic plane:
1) Galactic longitude l = 40$^\circ$ to l = 100$^\circ$ and
2) Galactic longitude l= 140$^\circ$ to l = 200$^\circ$ 
\cite{Sinnis,Sinnis_update}.
Both region encompass Galactic latitudes $|{\rm b}| < 5^\circ$.
In the first region (inner Galaxy),
a detection of a signal with a statistical significance of
4.5\,standard deviations is reported, corresponding to an
integral flux of
$\Phi (E > 3.5\,{\rm TeV})\, =\, 6.4\pm 1.4\pm 2.1 \times 10^{-11}\,$
photons\,cm$^{-2}$\,s$^{-1}$\,sr$^{-1}$.
This result is consistent with an extrapolation from the 
1-30 GeV flux measured by EGRET to 3.5\,TeV using a differential
spectral index of approximately $\alpha \sim 2.6$.
The observed flux could be due to both unresolved point sources and
true diffuse emission from the Galactic plane.
In the second region (outer Galaxy), however,
no evidence for a signal is obtained 
by Milagro and a limit on the
integral $\gamma$-ray flux is obtained
$\Phi (E > 3.5\,{\rm TeV})\, <\, 5.0 \times 10^{-11}\,$
photons\,cm$^{-2}$\,s$^{-1}$\,sr$^{-1}$
(99\% CL).
The limit is not a significant discrepancy with the
extrapolated EGRET spectrum as one expects a lower flux in the
outer region of the Galaxy.

The Tibet air shower array reported improved limits on the
diffuse Galactic $\gamma$-ray flux at energies
of 3 and 10\,TeV \cite{Ohnishi}.
These limits are not in disagreement with the results from
Milagro -- they lie just above the extrapolation 
of the EGRET flux using a differential spectral index of 2.6.

\subsection{Galactic Plane -- Selected Regions}
Results from observations of
specific regions of the Galactic plane were reported by a number
of instruments.
Both Milagro and HESS reported noteworthy detections of
diffuse emission at TeV $\gamma$-ray energies.

As an instrument with a wide-FOV, Milagro is well-suited to
making a survey for $\gamma$-ray emission over the entire
overhead sky.
The search for extended sources is an extension of the general
survey, and in this search there were two regions that showed
strong evidence for VHE $\gamma$-rays: the Cygnus region near
the Galactic plane and the Crab Nebula region.
A more detailed study of the Cygnus region was made
using an angular square bin
size that varied from 2.1$^\circ$ to 5.9$^\circ$ \cite{Smith2}.
For the largest bin size of 5.9$^\circ$, a diffuse source of
VHE $\gamma$-rays was detected in the Cygnus region with
a pre-trials statistical significance of 6.7 standard deviations.
The authors estimate the probability for a
chance detection of this excess due to a statistical fluctuation to
be $1.0 \times 10^{-5}$ \cite{Smith2_update}.
The excess appears to be spatially correlated with the diffuse
$\gamma$-ray emission detected by EGRET at GeV energies in this
region, and approximately 5$^\circ$ away
from the unidentified HEGRA source TeV 2032+4130.
Figure~2 shows the significance map resulting from the
Milagro observations.

\begin{figure}[h]
\begin{center}
\includegraphics*[width=0.65\textwidth,angle=0,clip]{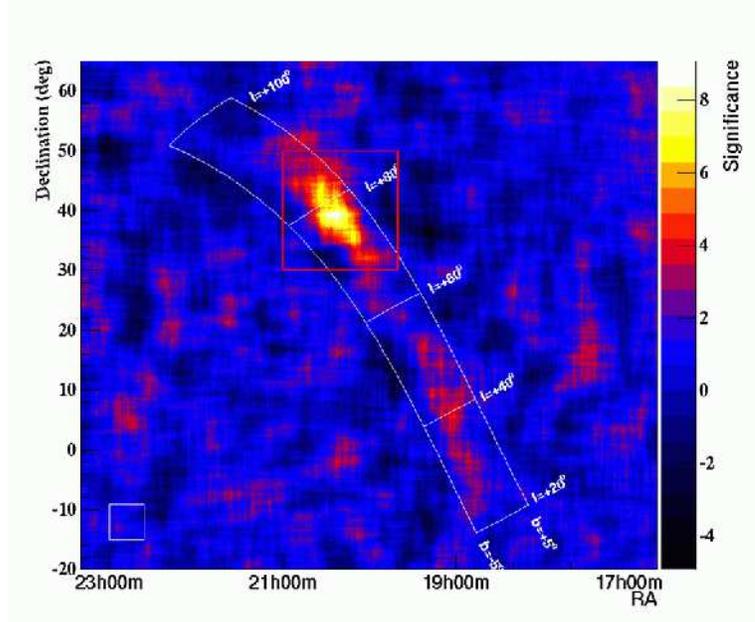}
\caption{\label {fig2} 
Significance map of Milagro extended source survey in the
Cygnus arm region \cite{Smith2}.
The color scale shows the excess event 
significance as a function of right
ascension (horizontal axis) and declination (vertical axis).
Data are binned in 5.9$^\circ$ bins, so neighboring points are
highly correlated.
The white boxes indicate ranges of Galactic coordinates.
The red box is 20$^\circ$ x 20$^\circ$ region around the bin with the
greatest significance.}
\end{center}
\end{figure}

The Galactic center is probably the
richest region that we can study
in terms of potential VHE sources, both 
point and diffuse sources.
CO observations reveal approximately 50 million solar
masses of molecular clouds in the central 300 parsecs
of the Galaxy.
HESS reported on a detailed study of the Galactic center 
region using 50\,hrs of data taken in 2005 
\cite{Hinton,Hinton_update}.
As discussed earlier, they detect two strong points sources in this
region (HESS J1745-290 and HESS J1747-281, see Table~1).
After subtracting the estimated signal from these two sources,
HESS reports evidence for diffuse $\gamma$-ray emission along the plane,
with a pre-trials statistical significance of 14.6 standard deviations.
An important feature of the diffuse excess is that it correlates
reasonably well with the molecular material as revealed by
sub-millimeter observations.
Thus, although there certainly may be additional point sources
that are not yet resolved in this study, the data provide strong
evidence for the production of $\gamma$-rays from interactions of
cosmic rays with molecular material.

Limits on diffuse $\gamma$-ray emission from selected regions
were reported by CANGAROO-III, who observed two regions
along the Galactic plane at at energy of 600 GeV \cite{Ohishi},
and by
Whipple, who observed a region in the Cygnus arm of the Galaxy,
near to the Milagro excess \cite{Atkins}.
Neither limit is inconsistent with the reports discussed above.
In the case of the Cygnus region, the Whipple pointing is
not coincident with the peak of the Milagro excess and that
excess is spread out of a relatively large angular extent.
Weak evidence for a diffuse $\gamma$-ray signal was presented
by TACTIC from an observation in the Crab region \cite{Bhat}.

\subsection{Other Diffuse Sources}

A variety of other sources of VHE diffuse $\gamma$-radiation have
been postulated, including starburst galaxies and nearby
galaxies that may have a large component of dark matter.
Previously, the CANGAROO telescope reported the detection of
TeV $\gamma$-rays from the starburst galaxy NGC 253 \cite{Itoh}.
New observations of NGC 253 have been made by CANGAROO-III,
but no results were reported.
However, HESS reported results from recent observations 
that are clearly inconsistent with the original CANGAROO detection.
Using data taken in 2003, HESS finds no evidence for $\gamma$-ray
emission from NGC 253 and sets an upper limit on the integral flux
of  $\Phi (E > 300\,{\rm GeV})\, <\, 1.9 \times 10^{-12}\,$
photons\,cm$^{-2}$\,s$^{-1}$ (99\% CL), assuming a point-like source
and an upper limit of
$\Phi (E > 300\,{\rm GeV})\, <\, 6.3 \times 10^{-12}\,$
photons\,cm$^{-2}$\,s$^{-1}$ (99\% CL), assuming a source
size of 0.5$^\circ$ \cite{Lemoine,Lemoine_update}.
Thus, the present status of NGC 253 as a VHE $\gamma$-ray emitter
is very uncertain.
A model for diffuse emission from NGC 253
was presented that argued that the VHE $\gamma$-ray
flux would be hard to detect by atmospheric Cherenkov telescopes,
but would possibly be detectable by the future space telescope
GLAST \cite{Domingo}.

Dark matter in the form of supersymmetric neutralinos could
annihilate to produce high-energy gamma rays.
Whipple reported a limit on the $\gamma$-ray emission from
the dwarf galaxies Draco and Ursa Minor and from the galaxies
M32 and M33 \cite{Hall}.
The Tibet air shower array reported flux upper limits from 
observations of M31, M32, and M87, interpreted in the
context of dark matter annihilation \cite{Zhang}.
The possibility of a dark matter signal in the detected
$\gamma$-ray flux from the Galactic Center region is discussed
in the following section.

Perhaps the most scientifically interesting diffuse radiation is
an isotropic radiation that could result from many unresolved 
extragalactic point
sources or from the injection of very high energy radiation
from processes in the early universe.
EGRET has measured the isotropic diffuse radiation at energies
from 30 MeV to 100 GeV \cite{Sreekumar}.
New measurements of this radiation in the GeV (and possibly TeV)
energy bands will likely have to wait until the launch
of GLAST.
Papers describing models for the extragalactic diffuse emission
were presented \cite{Kneiske,Grenier}.

\section{OG 2.2: Galactic Sources}

Understanding the origin of the cosmic rays is one
of the most outstanding questions in all of astrophysics.
The cosmic-ray spectrum exhibits an unbroken power-law form
over an enormous range of energies, from
$10^9$\,eV to $10^{20}$\,eV, and the energy density of the
cosmic rays in our Galaxy is $\sim 1\,$eV\,cm$^{-3}$, 
comparable to that in the cosmic microwave background.
The fact that the bulk of the cosmic ray are charged,
and that the Galaxy has a several $\mu$G irregular magnetic field, 
has hampered our ability to deduce their origin since their
directions are scrambled on their way to us.
Thus, VHE $\gamma$-rays are the most direct probe of extreme,
non-thermal astrophysical sources that could be the production
sites of the cosmic rays.

Conventional wisdom holds that the bulk of the cosmic rays up to
an energy of 10$^{14}$\,eV are produced in supernova remnants (SNRs)
in our Galaxy.
This wisdom comes in part from the issue of energetics --
SNRs are perhaps the only Galactic source with sufficient luminosity
to power and replenish the cosmic rays -- and in part from
the fact that we see strong evidence in X-ray data
for non-thermal acceleration of particles to TeV energies in SNR shocks.
By 2003, evidence for VHE $\gamma$-ray emission from a small number
of SNRs had been presented,
but the significances for these detections were marginal.
At this meeting, strong detections of numerous SNRs were reported
by several instruments, most notably HESS, and it now appears
unambiguous that SNRs are important sources of
VHE $\gamma$-radiation.
A very significant step forward has been taken 
towards demonstrating that SNRs are in fact the primary sites of
high-energy cosmic ray production.

Other potential sources of VHE $\gamma$-rays in the Galaxy
include pulsars and their associated nebulae (pulsar wind nebulae, PWN),
binary star systems (such as microquasars and binary pulsars), 
and OB stellar associations.
All-sky surveys at TeV energies by the Milagro and Tibet
air-shower arrays have revealed that there is only one other
steady source in the Galaxy with a $\gamma$-ray flux comparable to
the Crab Nebula (the extended source in the Cygnus region, see
Section~4.2), and hence when searching for new Galactic sources,
it is critical to achieve a flux sensitivity well below
the level of the Crab.
The newly commissioned atmospheric Cherenkov telescopes
(CANGAROO, HESS, MAGIC, and VERITAS) are all intended to reach this
sensitivity.

\subsection{Galactic Sources: Old \& New}
At ICRC 2003, the sky map of Galactic VHE sources contained
eight sources that were reported with reasonable confidence \cite{Grading}.
These sources include three PWN (Crab Nebula, PSR 1706-44, and Vela Pulsar),
four SNRs (SN 1006, RX J1713-3946, Cas A, and RX J0852-4622)
and one unidentified source in the Cygnus region (TeV 2032+413).
There was also a hint of something interesting at the Galactic center.

At this meeting, the situation with Galactic sources has changed dramatically
as a result of the HESS observations and new detections.
In addition to carrying out the Galactic plane survey (Section 3.1), 
HESS has targeted numerous Galactic sources of interest.
HESS reports the detection of 15 new sources that are likely to have
Galactic origins.
In addition to the 11 new sources listed in Table~2
(RX J1713-3946 had been reported earlier), HESS has detected
the composite SNR/PWN MSH 15-52, the binary pulsar PSR B1259-63, 
an unidentified object HESS J1303-631 (in the same field of view as
PSR B1259-63), and a new source in the Vela region 
HESS J0835-456 (Vela-X).
There is also the confirmation of the Galactic center by MAGIC.

HESS also presented the non-detection of three sources: 
PSR 1706-44, Vela Pulsar, and SN 1006.
For PSR 1706-44, the HESS flux upper limits are below 
the earlier measurements reported by the CANGAROO and Durham
telescopes.
For SN 1006, the HESS flux upper limits are below the earlier
measurements reported by CANGAROO and HEGRA CT1 \cite{Mori}.
Thus, the validity of both objects as probable VHE $\gamma$-ray
sources is questioned and they are removed from the Galactic source list.
For Vela, the HESS source J0835-456 is near the X-ray PWN, but no significant
emission is seen at the Vela Pulsar position \cite{Khelifi}.
Thus, this source is thus named Vela-X to distinguish it from the pulsar
\cite{HESS_VelaX}.
Upper limits on these three sources were also presented by CANGAROO
\cite{Tanimori,Tanimori_update}.

Thus, in summary, fifteen new Galactic sources were presented at ICRC 2005, but
three sources claimed earlier were cast in doubt.
The current table of VHE Galactic sources now has 20 objects, as 
shown in Table~3.

\begin{table}   
\caption{\label{galactic-sources} 
Galactic VHE sources.
A comparison between the VHE Galactic sources established in 2003 and
in 2005 is made, showing the dramatic improvement in the source count.
The sources are divided by source type; some sources have a tentative 
association.
In addition to the sources listed here, one must include the
sources with tentative or unknown associations from the HESS Galactic
plane survey (i.e. the eight sources below the line in Table~2).
There are thus 20 Galactic 
sources in the current
VHE $\gamma$-ray source catalog.}
\vspace{0.3cm}
\begin{center}
\begin{tabular}{||l|l|l||} \hline \hline
Source Type & 2003 Sources & 2005 Sources \\
\hline
& & \\
Pulsar Wind Nebulae (PWN) & Crab Nebula & Crab Nebula \\
& PSR 1706-44 & \\
& Vela Pulsar & \\
& & SNR G0.9+0.1 \\
& & MSH 15-52 \\
& & HESS J0835-456 (Vela-X) \\
& & \\
Supernova Remnants (SNRs) & SN 1006 & \\
& RX J1713-3946 & RX J1713-3946 \\
& RX J0852-4622 & RX J0852-4622 (Vela Jr) \\
& Cas A & Cas A \\
& & \\
Galactic Center & & SGR A \\
$\mu$-Quasar & & LS 5039 \\
Binary Pulsar & & PSR B1259-63 \\
Unidentified & TeV 2032+413 & TeV 2032+413 \\
& & HESS J1303-631 \\
& & \\
\hline
\end{tabular}
\end{center}
\end{table}

\subsection{Pulsar Wind Nebulae and Pulsars}

The Crab Nebula was the first source detected at TeV $\gamma$-ray energies
and is the standard candle for the field.
New measurements of the Crab Nebula were reported by a number of instruments
at this meeting.
HESS reported results from the very strong detection of the Crab Nebula and
determined a best-fit centroid location for the VHE emission of  
($\alpha$,$\delta$) = (05h 34m 8s, 22h 1m 11s) with a statistical uncertainty
of 5 arc-sec and a systematic uncertainty of 20 arc-sec \cite{Masterson}.
This centroid is consistent with the known positions of the pulsar and the
X-ray PWN.  The VHE emission is consistent with originating from a point 
source;
HESS placed an upper limit on the extent of the source of $< 2$\,arc-min 
(99\% CL).
MAGIC reported results from data taken on the Crab Nebula in 2004 and 2005
\cite{Wagner}; MAGIC measured the energy spectrum from 100\,GeV to 6\,TeV 
and is
developing methods to push their analysis energy threshold below 100\,GeV.
A fit to the MAGIC data between 300 and 3000\,GeV yields a differential 
spectral
index of $\alpha = 2.58\pm 0.16$.
STACEE reported on measurements of the Crab Nebula between 100 and 1500\,GeV
using a new technique for $\gamma$/hadron separation related to the smoothness
of the Cherenkov shower front \cite{Kildea2}.  
Other studies of the Crab were reported by CANGAROO (at large zenith angles) 
\cite{Nakamori} and TACTIC \cite{Dhar}.

HESS reported the detection of five new sources that are most likely associated
with PWN: SNR G0.9+0.1, MSH 15-52, HESS J0835-456 (Vela-X), HESS J1616-508,
and HESS J1825-137, and the non-detection of PSR 1706-44 and the Vela Pulsar
(see Tables 2 and 3). 
SNR G0.9+0.1 was detected in the same field of view as the Galactic
Center \cite{Khelifi}; this source is a composite SNR, but the VHE emission 
detected
by HESS is consistent with the PWN position and not with the SNR shell.
The source is one of the weakest detected (flux $\sim 2$\% Crab) and
the energy spectrum measured by HESS is consistent with a single power-law form
with differential spectral index of $\alpha = 2.40 \pm 0.11 \pm 0.20$.
MSH 15-52 is also a composite SNR, containing a remnant, a 150\,ms pulsar,
and a pulsar wind nebula.
The VHE $\gamma$-ray emission detected by HESS is clearly extended 
along the NE-SW
direction \cite{Khelifi}, 
in a pattern that is consistent with the X-ray morphology detected by ROSAT.
This is the first evidence for an extended PWN source at TeV $\gamma$-ray 
energies.
The HESS energy spectrum for MSH 15-52 (flux $\sim 15$\% Crab) is fit by
a single power-law shape from 280\,GeV all the way up to 40\,TeV, with 
differential spectral index of $\alpha = 2.27 \pm 0.03 \pm 0.20$.
The Vela region is rich in potential VHE sources.
HESS detected a relatively strong (flux $\sim 50$\% Crab) source, 
HESS J0835-456
(Vela-X),
that is located on the south side of the X-ray PWN 
\cite{Khelifi,HESS_VelaX}.
The morphology of Vela-X appears to be extended and the VHE energy spectrum
exhibits clear downward curvature above 1\,TeV.
HESS J1825-137 was discovered during the survey of the Galactic plane and
the VHE emission is plausibly consistent with originating from the 
PWN G18.0-0.7.
Like Vela-X, J1825-137 has extended emission that is positionally
coincident with one side of the PWN.  This may result from the 
remnant expanding
into a inhomogeneous medium.
Theoretical modeling to understand the morphology of the PWN and to explain the
relatively large size of the PWN was presented by de Jager 
\cite{deJager,deJager_update}.

Although there is clear evidence that the nebulae associated with pulsar winds
produce VHE $\gamma$-rays, there have been no detections of pulsed emission
at these energies that would come from the pulsar itself.
EGRET detected eight pulsars in the GeV range, but
the origin of the high-energy
pulsed emission in pulsars is still mysterious; 
the various models (polar cap, slot gap, outer gap, etc.) generally 
predict cut-offs in the pulsed spectrum in the 1-100\,GeV $\gamma$-ray range.
New upper limits on the flux of VHE emission from the Crab Pulsar were
reported by HESS \cite{Konopelko}, MAGIC \cite{Lopez}, PACT \cite{Singh},
and STACEE \cite{Kildea2}.
The lowest energy points came from
MAGIC's limits at the two energies of 90 and 150\,GeV, slightly above
the energy range explored in an earlier measurement by CELESTE \cite{CELESTE}.
HESS also reported upper limits on the VHE emission from the Vela Pulsar
and PSR 1706-44.
The SHALON \cite{Sinitsyna1} and PACT \cite{Vishwanath1} telescopes
presented results from observations of the Geminga Pulsar, but VHE $\gamma$-ray
emission from this source is unconfirmed at the present time.
The class of rapidly spinning millisecond pulsars are not known to emit
$\gamma$-rays, but these binary systems are certainly potential targets
for VHE telescopes.
MAGIC reported results from observations of two millisecond pulsar systems
PSR B1957+20 and PSR J0218+4232 \cite{Ona-Wilhelmi};
no pulsed or continuous emission was detected from either source and upper
limits on the steady VHE emission and the pulsed fraction at energies
above 115\,GeV were presented.

\subsection{Supernova Remnants}

As discussed earlier, prior to this year we had evidence of
VHE $\gamma$-ray emission from four supernova remnants (SNRs):
SN 1006, RX J1713-3946, RX J0852-4622, and Cassiopeia A (Cas A).
HESS and CANGAROO have confirmed emission from RX J1713-3946
and RX J0852-4622 and have been unable to detect SN 1006
\cite{Rowell,Tanimori}.
HESS has also detected three sources in their
survey of the Galactic plane that can be plausibly associated
with supernova remnants:  HESS J1640-465, HESS J1813-178, and
HESS H1834-087 (see Table~2).
We can now say with confidence that shell-type SNRs are general
sources of VHE $\gamma$-rays.

\begin{figure}[ht]
\begin{center}
\includegraphics*[width=0.5\textwidth,angle=0,clip]{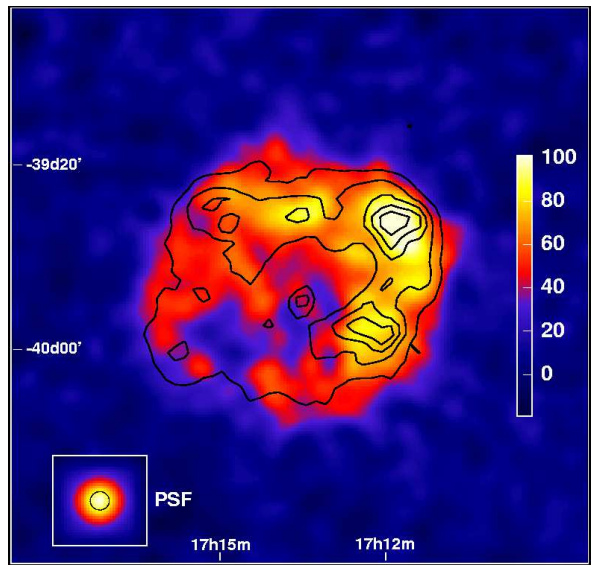}
\caption{\label {fig3} 
VHE $\gamma$-ray excess map of the SNR RX J1713-3946 from HESS
\cite{Berge,Berge_update}.
The black contour lines show the surface brightness in 1-3\,keV
X-rays as measured by ASCA.}
\end{center}
\end{figure}

RXJ 1713-3946 is a large ($\sim 1^\circ$) SNR that has been well studied by
a number of X-ray instruments (ROSAT, ASCA, Chandra, and XMM-Newton).
The source was detected earlier by CANGAROO and by HESS using
two telescopes in the construction phase. 
The current measurements by HESS are part of a very strong detection
of the source based on 40\,hrs of data
taken with the full four-telescope array \cite{Berge}.
As shown in Figure~3, HESS well reconstructs the 
spatial morphology of the $\gamma$-ray emission; the SNR shell has been
clearly resolved and the VHE emission strongly maps the pattern
seen in X-rays.
Even more, by dividing the data into three bands of energy
($E < 0.6\,$TeV, 0.6\,TeV$ < E < 1.4$\,TeV, and $E > 1.4$\,TeV)
HESS is able to show that 
the VHE morphology does not change appreciably with energy.
The energy spectrum is well reconstructed from 200\,GeV to 30\,TeV;
the spectral index is hard ($\alpha \sim 2.2$), but the spectrum
exhibits curvature and does not well fit a single power-law form.
The quality of the HESS data is sufficient to allow spatially-resolved
spectral determination - the spectrum measured in 14 different
regions of the remnant are not found to vary significantly.

RX J0852-4622 (Vela Jr) is a large ($\sim 2^\circ$) shell-type
supernova remnant discovered in the ROSAT all-sky survey and
exhibiting non-thermal X-ray emission.
The previous CANGAROO detection found evidence for VHE
$\gamma$-ray emission from one part (NW rim) 
of the SNR shell.
New data from both CANGAROO \cite{Tanimori} and HESS \cite{Komin}
confirm that this source is indeed a relatively strong VHE
$\gamma$-ray emitter.
The HESS data show a VHE spatial morphology that is large
and extended and is well correlated with the X-ray morphology
as seen by ROSAT at 1-3\,keV.
HESS measures a relatively hard spectrum with a differential
spectral index $\alpha \sim 2.1$.
The CANGAROO measurements indicate a somewhat small spatial
extent and a somewhat softer energy spectrum.
However, the differences 
between the two experiments are probably not significant at this point.
Some uncertainty remains regarding the dominant mechanism behind
the VHE $\gamma$-ray emission; both electron (inverse-Compton) and
proton ($\pi^o$) models seem plausible \cite{Komin}, and more detailed
modeling is needed to differentiate between these two scenarios.
Regardless, RX J0852-4622 is the second shell-type SNR to be
detected and well-imaged at VHE $\gamma$-ray energies.

In addition to the detections listed already, HESS
has carried out
an extensive campaign to observe a number of other promising
supernova remnants that may be sources of high-energy cosmic rays.
Initial results were reported on data taken on five SNRs:
W28 (G6.4-0.1), SN 1987A, IC 443 (G189.1+3.0),
Monoceros Loop (G205.5+0.5), RCW 86 (G315.4-2.3) \cite{Rowell}.
No strong VHE $\gamma$-ray emission was found from
any of these sources, although there was some evidence of emission from
W28.
At higher energies, the Tibet air shower array presented upper
limits on PeV $\gamma$-ray emission from the large Monogem Ring using
data taken over a seven year period \cite{Amenomori2}.
The limits are well below the claimed detection by the MAKET-ANI
experiment \cite{Chilingarian}.
There were also reported observations of the remnants G40.5-0.5 by
the Tibet air shower array \cite{Amenomori3} and Tycho by
the SHALON telescope \cite{Sinitsyna1}

The numerous X-ray and VHE $\gamma$-ray measurements clearly show
that supernova remnants accelerate particles to energies of 50\,TeV
and possibly higher. 
The question remains what fraction of the energetic particles are electrons
and what fraction are protons.
Detailed diffusive shock acceleration models have been developed
to explain the broad-band emission from SNRs.
A non-linear model for SN 1006 points out the key importance
of the density of hydrogen gas that serves as the target material;
the HESS non-detection argues for $N_H < 0.1$\,cm$^{-3}$ \cite{Ksenofontov}.
Another study points out that cosmic-ray ions may also be efficiently
accelerated in SNRs and could possibly modify the hydrodynamics of the 
remnant and require very compressed magnetic fields \cite{Ellison1}.
The general issue of whether SNRs are the origin of the high-energy
cosmic rays was considered in the context of the general non-linear
kinetic theory for acceleration \cite{Berezhko}.
The evidence for strong magnetic field amplification in young
SNRs comes from the observation of steep synchrotron spectra and
sharp features observed in X-rays.
Thus, electrons could in principle be the source of the VHE $\gamma$-ray
emission, but this scenario would required magnetic fields that are
generally not supported by the X-ray data or the models.

\subsection{Galactic Center}

Understanding the Galactic center region is an important
scientific quest.  Not only do we wish to understand
the center of the galaxy that we live in, but in
learning about the Milky Way we could very well derive important
important regarding normal, spiral galaxies like our own.
The Galactic center has been a promising target for 
$\gamma$-ray telescopes for some time.
EGRET detected strong emission from the general region in
the MeV/GeV band \cite{EGRET-GC}, however the angular resolution of EGRET
(and the presence of an ubiquitous glow of diffuse radiation)
made a clear interpretation of the nature of this emission
difficult.

Previous evidence for VHE $\gamma$-ray emission from the Galactic center
came from the CANGAROO and Whipple telescopes.
At ICRC 2005, both HESS \cite{Rolland} and MAGIC 
\cite{Bartko,Bartko_update} 
reported detections with
substantially improved significance and resolution (spatially and
spectrally).  HESS reported results from observations made in 2003 and
2004 encompassing over 50\,hrs of live time.
The detection significance was very high ($> 40\sigma$) and
HESS was able to map out the position and extent of the $\gamma$-ray
emission with excellent precision.
With the assumption that the emission came from a point source, HESS
localized the emission centroid to within $6 \pm 10 \pm 20$ arc-secs
of Sagitarius A* (SGR A*, the putative black hole).
However, the HESS data show that the emission has an extension of
$1.9 \pm 0.2$ arc-mins in addition to a point-source like core.
A map showing the positions and error boxes of the various 
$\gamma$-ray observations of the Galactic center is shown in Figure~4.
the HESS emission is compatible with an INTEGRAL source, with a
supernova remnant (SGR A East), and with the black hole, SGR A*, but
is not compatible with a nearby unidentified EGRET source.
HESS measured a power-law spectrum that is well fit by a single,
unbroken power-law form between 160\,GeV and 10\,TeV with a
differential spectral index $\alpha = 2.20 \pm 0.09 \pm 0.15$.
The light curve measured by HESS for the five month observation
period in 2004 was flat, with no evidence for variability on
time scales as short as 10\,min. 
MAGIC reported results from observations made in 2005 encompassing
15\,hrs of live time.  MAGIC employed the large zenith angle (LZA)
technique as the Galactic
center culminates at an elevation of 32$^\circ$ at La Palma.
This lead to an elevated energy threshold of $\sim 600$\, GeV
for the observations.
The centroid of the MAGIC $\gamma$-ray excess is consistent with
the position of SGR A* and with a point source hypothesis, given
the statistical significance ($\sim 6\sigma$) of the signal.
The spectrum measured by MAGIC is hard with differential index
$\alpha = 2.3 \pm 0.4$ and in very good agreement with that
measured by HESS, as shown in Figure~4.

\begin{figure}[ht]
\begin{center}
\includegraphics*[width=0.95\textwidth,angle=0,clip]{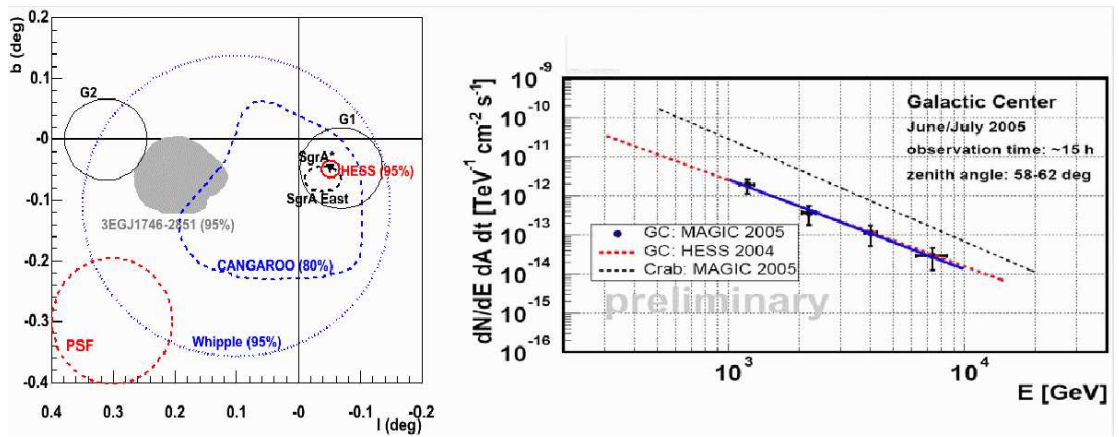}
\caption{\label {fig4}
Measurements of the VHE $\gamma$-ray source at the Galactic
center by HESS \cite{Rolland} and MAGIC \cite{Bartko,Bartko_update}.  
Left:  error boxes for the
source location as reported by HESS, CANGAROO, and Whipple,
as indicated.  Solid circles, G1 and G2, indicate INTEGRAL
sources and the shaded region indicates an unidentified
EGRET source.  The locations of Sgr A* (star) and
the SNR Sgr A East (dashed black line) are also shown.
Right: Differential energy spectrum for the source as 
measured by MAGIC in 2005.  Also shown are the MAGIC
spectrum for the Crab Nebula (dark green dot-dash) and
an approximation of the Galactic center spectrum measured
by HESS (red dot-dash).}
\end{center}
\end{figure}

Now that the Galactic center is a firmly established VHE source,
the question remains as to the nature of this strong, and
somewhat unexpected, emission
($\sim 50$\% of the Crab luminosity at very high energies).
Astrophysical possibilities 
include shock acceleration at the supernova
remnant SGR A East (or an obscured SNR or plerion), 
interactions of stellar winds or VHE cosmic rays with ambient
material, 
non-thermal processes associated with the black hole itself,
or something completely different!
There is also the tantalizing potential of new physics:
dark matter in the form of supersymmetric WIMPs could produce
VHE $\gamma$-rays through the neutralino annihilation process.
Some discussion of the viability of the dark matter hypothesis
was presented at ICRC 2005 \cite{DM}; no significant constraints
on dark matter models can be made at the present time for
a variety of reasons.
Foremost is the fact that the properties of the VHE
emission detected so far are completely compatible with an
astrophysical origin, and the region will need to be studied 
in considerable more detail in order to disentangle the astrophysical
contributions to the $\gamma$-ray signal. 
It is perhaps ironic that in the presence of such a strong source
of TeV emission, the Galactic center may in fact be a very difficult
region in which to search for dark matter.
Other difficulties arise because of the substantial
uncertainties in the actual dark matter profile at the core
of the Galaxy and in the model parameters for the supersymmetric
extension to the Standard Model.

\subsection{Other Galactic Sources}

As shown in Table~3, three additional types of Galactic sources that
have been detected at VHE $\gamma$-ray energies are a 
microquasar, a binary pulsar, and several unidentified objects.

Microquasars generally exhibit strong emission across a broad
range of wavelengths, with jets observed
in radio and rapid variability in X-rays.
The standard picture of a microquasar is a binary system in which
a normal star orbits around a compact object.
Mass lost from the star falls into an accretion disk where it 
can be heated and ejected or can fall into the compact object.
In some ways, microquasars can be considered active galaxies in
miniature, and so there is the hope that  by understanding microquasars
we can shed light on the mechanisms that power AGN.
HESS has observed a number of microquasar targets and reported
flux upper limits for four sources,
GRS 1915+105 (reported earlier by HEGRA),V4641 SGR, GX 339-4 and Circinus X-1
\cite{Nolan}, and detected one source, LS 5039 from data
taken as part of the Galactic survey \cite{deNaurois,deNaurois_update}.
The nature of the compact object (i.e. neutron star or black hole)
in LS 5039 is not clear, and, although the source is relatively dim in
X-rays, the fact that it has a 
massive (20\,M$_\odot$) companion star may have aided in its detection
in VHE $\gamma$-rays.
The HESS data, consisting of $\sim 10$\,hrs of livetime, show
a clear detection of a point source whose position is consistent
with the known location of LS 5039 and is not consistent with a
nearby SNR and pulsar.
The HESS source is also consistent with the large error box of
the EGRET source 3EG J1824-1514.
HESS measured a relatively hard spectrum at very high energies
(spectral index $\alpha \sim 2.1$) and
although source variability might be anticipated in a microquasar,
none has so far been detected from the HESS observations.
The establishment of this new source class is an important achievement;
more extensive observations of such sources will tell us
how prevalent microquasars are at very high energies and whether
they provide clues relating to
the acceleration processes in active galactic nuclei.

The binary system PSR B1259-63 consists of a 48\,ms radio pulsar that
orbits a bright and massive (10\,M$_\odot$) Be star.
Radio and optical observations indicate that the Be companion has a dense
stellar disk, presumably formed from mass outflow.
The 3.4\,yr orbit of the pulsar is highly eccentric, causing the pulsar
to pass within $\sim 20$ star radii of the companion at periastron.
HESS reported on observations taken in early 2004 , before and after
periastron, followed by observations made in 2005 
\cite{Schlenker,Schlenker_update}. 
The source was well detected in the first observation period, but
not in the second.
In the first period, the detected $\gamma$-ray flux was consistent
with coming from a point source and the spectrum was consistent
with a single power-law form (spectral index $\alpha \sim 2.7$), but
the flux level was highly variable, on time scales of days,
as shown in Figure~5.
The HESS results thus reveal a new type of VHE $\gamma$-ray source and
the first variable galactic source discovered at these energies.
A general model, that successfully predicted the detection of
VHE $\gamma$-rays, postulates that the pulsar produces a relativistic
wind of electrons with energies exceeding 10\,TeV \cite{Kirk}.
These
electrons upscatter optical/ultraviolet photons from the Be star via the
inverse-Compton effect.
The TeV emission is expected to be variable as the pulsar wind approaches,
and then retreats, from periastron.
Although the inverse-Compton scenario qualitatively describes
the VHE $\gamma$-ray light curve and the multi-wavelength spectral
energy distribution (SED), more measurements are required to fully
understand this complex system.

\begin{figure}[ht]
\begin{center}
\includegraphics*[width=0.90\textwidth,angle=0,clip]{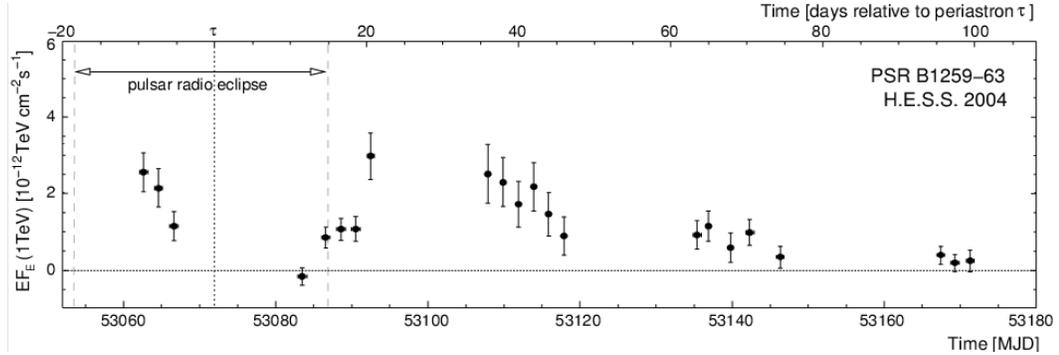}
\caption{\label {fig5} 
VHE $\gamma$-ray light curve of the binary pulsar PSR B1259-63
as measured by HESS in 2004 \cite{Schlenker,Schlenker_update}.
The integral $\gamma$-ray flux above 1\,TeV is plotted as
a function of time (in days) relative to periastron.
This result marks the first variable Galactic source to be
detected at VHE $\gamma$-ray energies.}
\end{center}
\end{figure}

A key goal of any wave band of astronomy is to discover
new astrophysical sources and new phenomena
not yet seen at other wavelengths.
The gamma-ray burst (GRB) is the archetypical example of a source
first detected at $\gamma$-ray energies that would have been very
difficult to discover at lower energies.
The detection of the first unidentified VHE $\gamma$-ray source,
TeV 2032+413, by HEGRA thus marked an important milestone for the
field \cite{Horns1}.
HESS has reported the detections of four additional sources that
are not yet identified, or are poorly identified, with objects
at other wavebands:
HESS J1614-518, HESS J1804-216, HESS J1837-069, and HESS J1303-631.
The first three sources were discovered as part of the Galactic
plane survey \cite{Funk} and are listed in Table~2.
The last source, HESS J1303-631, was discovered serendipitously
in the same field as the binary pulsar PSR B1259-63 \cite{Raue}.
The source appears to be spatially extended and constant in flux
during both the 2004 and 2005 observing periods.
No clear counterpart has been identified (e.g. there is no 
significant X-ray source within the error circle of the
HESS detection) \cite{Raue_update}.
Additional VHE and multi-wavelength observations may clarify
the nature of this source, but it is exciting to speculate
that this source may represent a new class of ``dark'' TeV
$\gamma$-ray emitters.
Two other sources, discussed earlier, that do not have a
firm or plausible association are the Galactic center and
diffuse source in the Cygnus region reported by 
Milagro \cite{Smith2}.

\section{OG 2.3: Extragalactic Sources}

Extragalactic sources have been key components of the
observing program for VHE $\gamma$-ray telescopes for
many years.
In fact, it is fair to say that until 2004, most of the
exciting developments in the field came from observations
of extragalactic sources, especially active galactic nuclei (AGN)
of the blazar variety.

Blazars (quasars and BL Lacertae, or BL Lac, objects) are
important objects that have strong and variable emission at
most wavelengths where they are detected.
Relativistic jets can be seen, or inferred, in many blazars and
the sources often exhibit optical polarization and superluminal
motion.
The general model for a blazar is one involving a supermassive
($10^6 - 10^9$\,M$_\odot$) black hole surrounded by an
accretion disk. Matter falling towards the black hole
powers the hot accretion disk and perpendicular, highly
collimated jets.
Blazars are thought to be those AGN whose jets are aligned
towards the direction of Earth,
and the VHE $\gamma$-rays are presumably produced from
acceleration processes involving protons and electrons
in the jets.
The nature of jets (how they form, their composition,
their zones of emission, etc.) are a key astrophysical puzzle.
In particular, a key question is
whether the dominant beam particles
are electrons, that produce X-rays via synchrotron radiation
and TeV $\gamma$-rays via inverse-Compton processes,
or protons, that produce TeV $\gamma$-rays in 
cascades resulting from the interactions of protons with
ambient radiation fields or material.
Previous VHE observations have established that blazars
have strong and highly variable $\gamma$-ray emission,
that their emitted power is dominated by their high-energy
emission, that changes in 
the TeV emission are often directly correlated with changes
in the X-ray emission, and
that the sources generally have power-law
spectra extending out to 10\,TeV (with possible curvature
at the higher energies).

Other potential extragalactic 
VHE $\gamma$-ray sources include AGN
other than blazars (i.e. radio galaxies such as FR1 and
FR2 and radio-quiet spirals such as Seyferts),
gamma-ray bursts (GRBs),
galaxy clusters, dwarf galaxies, 
and starburst galaxies.
GRBs are discussed in Section~7 and the latter two are
discussed in Section~4.3.

An important consideration for extragalactic observations
is the potential impact on the detected $\gamma$-ray flux
from interaction with the extragalactic background light
(EBL).
The EBL is the total radiation in the
infrared(IR)/optical/ultraviolet(UV) bands from normal
star formation and radiation from dust, integrated
over the luminosity history of the universe.
Gamma rays in the 50 - 5000\,GeV range will interact
with EBL photons via the pair-production process.
Since
the EBL density is poorly known at the present time,
there is promise that spectral
measurements of extragalactic sources at a number of
redshifts could better determine the density and perhaps
constrain cosmological models that impact on the
evolution of the EBL.
Possible absorption effects in the 
VHE $\gamma$-ray spectra
from the sources Markarian 421 (Mrk 421), 
Markarian 501 (Mrk 501), and
H 1426+428 have been previously discussed, but a
general interpretation has not yet been established.

The VHE source catalog of extragalactic objects is
shown in Table~4.
There are currently 11 reasonably well-established
sources; ten of these are blazars and one is the
radio galaxy M87.
Four the blazars, PKS 2005-489, H 2356-309, 1ES 1218+304,
and 1ES 1101-232, were discovered in the last year and
presented at ICRC 2005. 
Three of these sources are the most distant objects
yet detected at very high energies.

\begin{table}   
\caption{\label{extragalactic-sources} 
Extragalactic VHE sources.
A comparison between the VHE extragalactic 
sources established in 2003 and
in 2005.
All blazars are of the BL Lac type.
The sources are ordered by their redshift values.
There are a total of 11 extragalactic sources in the current
VHE $\gamma$-ray catalog.}
\vspace{0.3cm}
\begin{center}
\begin{tabular}{||l|l|l|c||} \hline \hline
Source Type & 2003 Sources & 2005 Sources & Redshift \\
\hline
& & & \\
Starburst Galaxy & NGC 253 &      & 0.002 \\
Radio Galaxy     & M87     & M87  & 0.004 \\
Blazar & Markarian 421 & Markarian 421 & 0.031 \\
 & Markarian 501 & Markarian 501 & 0.034 \\
 & 1ES 2344+514 & 1ES 2344+514 & 0.044 \\
 & 1ES 1959+650 & 1ES 1959+650 & 0.047 \\
 & & PKS 2005-489 & 0.071 \\
 & PKS 2155-304 & PKS 2155-304 & 0.116 \\
 & H 1426+428 & H 1426+428 & 0.129 \\
 &            & H 2356-309 & 0.165 \\
 &            & 1ES 1218+304 & 0.182 \\
 &            & 1ES 1101-232 & 0.186 \\
& &  &\\
\hline
\end{tabular}
\end{center}
\end{table}

\subsection{New Extralactic Sources}

HESS reported the discovery three new AGN:
PKS 2005-489 \cite{Benbow},
H 2356-309 \cite{Pita},
and
1ES 1101-232 \cite{Tluczykont}.
PKS 2055-489 is a relatively nearby BL Lac of the
HBL (high frequency peaked) or extreme variety.
It was the target of previous observations by
the CANGAROO and Durham telescopes, but not
previously detected at VHE $\gamma$-ray energies.
HESS reported on data taken in 2003 (27\,hrs), during the
construction phase, and in 2004 (24\,hrs), with the full
four-telescope array.
No significant excess was found in the 2003 data, but
the 2004 data showed a statistical significance of
6.7$\sigma$, for an overall significance for both
data sets of 6.3$\sigma$.
The source is interesting in being very weak ($\sim 2.5$\% Crab
Nebula) and in having a very steep energy spectrum with a
differential spectral index $\alpha = 4.0 \pm 0.4$.
Given that other, more distant, AGN 
are detected with harder spectra, it is likely that the
steepness of the spectrum observed for PKS 2005-489
results from intrinsic properties of the source, as opposed
to absorption by the EBL.
No evidence for time variability was found on hourly, daily,
and monthly time scales.
X-ray data taken in October 2004 with XMM-Newton show a very
low state for the source and a very steep X-ray spectrum.
Thus, it is possible that this source was detected by
HESS in its low VHE state and that flaring and rapid variability
may be expected in the future.

H 2356-309 is an extreme BL Lac with an X-ray synchrotron
peak near 1.8\,keV.
Observations by HESS in 2004 totaling $\sim 39$\,hrs of live time
yielded a detection with a statistical significance of
9.0$\sigma$.
The energy spectrum is well fit by a single power-law form
with differential spectral index $\alpha = 3.09 \pm 0.16$,
as shown in Figure~6.
The spectrum is relatively hard at this redshift value
which may indicate
a lower amount of EBL absorption than had been suggested by
the earlier results from observations of
Mrk 421, Mrk 501, and H 1426+428. 
Some evidence was seen for monthly variability in the VHE
$\gamma$-ray flux.

1ES 1101-232 is also an extreme BL Lac with a flat X-ray
spectrum extending up to 100\,keV.
Observations by HESS in 2004 and 2005 totaling $\sim 40$\,hrs
of live time yielded a detection with a statistical significance
of 9.0$\sigma$, making this the most distant object yet
detected at very high energies.
No significant variability was detected on a variety of time
scales.
The energy spectrum for 1ES 1101-232 
by a single power-law form
with differential spectral index $\alpha = 2.88 \pm 0.17$.
As in the case of H 2356-309, this hard spectrum can the
interpreted as a constraint on the EBL, i.e. indicating
that the EBL density is lower than previously suspected.
The energy spectrum of 1ES 1101-232 as measured by HESS
is shown in Figure~6.

MAGIC reported the discovery of the BL Lac 1ES 1218+304
from 7\,hrs of data taken in 2005 as part of an extensive
observation campaign of AGN that targeted eight sources
\cite{Meyer}
The source was detected with a statistical significance
of 7.3$\sigma$ and with an energy spectrum well
fit by a single power law form with 
differential spectral index $\alpha = 3.3 \pm 0.4$
\cite{Bretz}.
No evidence for flux variability has been reported.
The energy spectrum of 1ES 1218+304 as measured by
MAGIC is shown in Figure~6.

\begin{figure}[ht]
\begin{center}
\includegraphics*[width=0.95\textwidth,angle=0,clip]{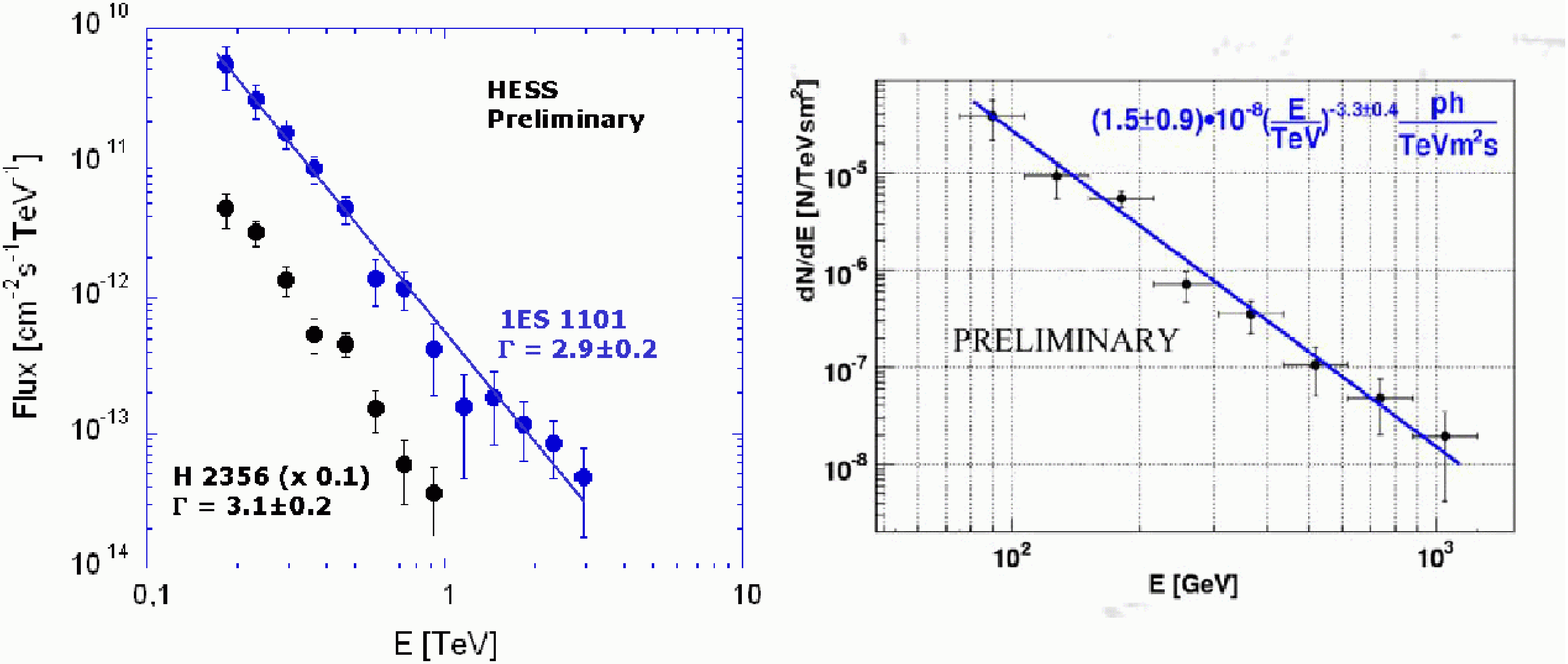}
\caption{\label {fig6}
Energy spectra for three new AGN as measured by HESS and MAGIC.
Left: HESS spectral measurements of the BL Lac sources H 2356-309 and
1ES 1101-232, with the indicated differential spectral indices
\cite{Hoffman}. The data for H 2356-309 have been scaled by a factor
of 0.1 for clarity.
Right: MAGIC measurement of the spectrum for the BL Lac source
1ES 1218+304, with the indicated differential flux determination
\cite{Bretz}.
Each of these three AGN, the most distant objects detected
at VHE $\gamma$-ray energies, can be fit by unbroken power laws to
$\gamma$-ray energies up to, and beyond, 1 TeV.}
\end{center}
\end{figure}

As shown in Figure~6, the spectra for the three most distant AGN
detected the VHE $\gamma$-ray energies can all be fit by single
power-law forms out to energies of 1\,TeV and beyond.
This observation can be used to constrain the density of the
EBL with some assumptions on its shape.
The HESS group argue that the AGN spectra constrain the EBL density
to be less than half of that expected from a typical model and
that the upper limit on the EBL density is approaching a stringent
lower limit set by counting resolved galaxies \cite{HESS-EBL}.
If the EBL density is significantly lower than expected,
we can concluded that
the universe is more transparent to VHE $\gamma$-rays than
was previously thought.

\subsection{Extragalactic Source Studies}

In addition to reports on the discovery of new AGN,
there were many contributions relating to detailed measurements
of known extragalactic sources.  Here we briefly summarize
these contributions and highlight a few of the most interesting
results.

The radio galaxy M87, at the center of the Virgo cluster, is
a unique object in being the closest, and the only non-blazar,
extragalactic source.
Originally discovered by HEGRA at VHE $\gamma$-ray energies,
M87 has been most recently studied by HESS and MAGIC.
Some models for M87 suggest that it may be similar to
a mis-aligned blazar.
It is also considered to be a likely nearby
source of ultrahigh energy cosmic rays.
The HESS observations of M87 encompass $\sim 31\,$hrs
of live time taken in 2003 and 2004 \cite{Beilicke}.
The source was detected with a statistical significance of
5.8$\sigma$ at an energy threshold of 380\,GeV, confirming
the original HEGRA detection.
The HESS $\gamma$-ray excess is consistent with coming from 
a point source that is close to the position reported by HEGRA
and is also close to the center of the galaxy.
There is evidence for flux variability in that the integral
flux measured by HESS is lower than that reported by
HEGRA, however more data taken by HESS in 2005 can very likely
solidify this possibility.
The MAGIC observations of M87 encompass $\sim 13$\,hrs of live time
taken in 2005 \cite{Albert}; no report of a detection was presented
at the meeting.

Markarian 421 (Mrk 421) is undoubtedly the best studied AGN
at VHE $\gamma$-ray energies due, in part, to its brightness and
its high degree of variability in both the X-ray and $\gamma$-ray
bands.
The source went into high states in both 2004 and 2005, 
whereupon it was studied by a number of VHE telescopes.
Figure~7 shows some of the results from a five-month monitoring
campaign by the Whipple telescope, along with X-ray data from
RXTE-PCA \cite{Grube}.
In April 2004, variations of approximately one order of magnitude
in size were observed in both the X-ray and VHE $\gamma$-ray
fluxes. Correlation between the X-ray and VHE flux levels was
observed, and the X-ray synchrotron peak shifted upward to above
2\,keV during the high state.
However, flares on hourly time scales were not resolvable.
Similar observations were carried out by HESS
\cite{Horns2} and by STACEE \cite{Carson}
during this same period.
The HESS observations were made a low zenith angle at energies
above 1\,TeV, where significant nightly variability was observed.
HESS reported an energy spectrum for Mrk 421 that has a clear
roll-over at high energies (above $\sim 5$\,TeV) and is best
fit by a power-law plus exponential form.
The STACEE data were taken from January to April 2004 at a median
$\gamma$-ray energy of 175\,GeV.
STACEE uses a new technique to extract the energy spectrum for
Mrk 421; the data are well fit by a single power-law form from
100\,GeV to 1.5\,TeV, with a measured differential spectral
index of $\alpha = 1.80 \pm 0.26$.
The spectral energy distribution (SED) for the STACEE data, in comparison
with the Whipple medium and high states, is shown in Figure~7.
The STACEE data show a flattening of the SED at low energies, which
might be expected if the peak of the inverse-Compton component
shifts to higher energies during a high flaring state.
The TACTIC instrument also made observations of Mrk 421 between
January and April 2004; the source was detected with a
statistical significance of 6.8$\sigma$ from $\sim 80$\,hrs of
live time \cite{Rannot}.  
The spectrum in the energy range between 2 and 9\,TeV can be fit
by a single power-law form for the TACTIC data.
Other results from observations of Mrk 421 in 2004 were reported
by CANGAROO, operating at low zenith angles \cite{Sakamoto}
and by PACT, where evidence for TeV $\gamma$-ray bursts was
presented \cite{Vishwanath2}.
PACT also reported on multi-wavelength
observations of Mrk 421 in 2003, where the source was not
detected \cite{Gupta}.

Results from recent observations of Mrk 421 were reported by MAGIC
for the time period between November 2004 and April 2005 \cite{Mazin1}
and by the first telescope of VERITAS during the first five months
of 2005 \cite{Cogan}.
In the MAGIC observations,
the source was strongly detected and hourly time variability was
clearly observed.
Correlation between the X-ray flux from RXTE-ASM and the 
$\gamma$-ray flux recorded by MAGIC was seen, but MAGIC did not see
any evidence for spectral hardening or for a flattening
in the energy spectrum at energies down to 100\,GeV.
The VERITAS data showed a significant detection of Mrk 421 and
Markarian 501 demonstrating that the telescope design is sound.
Some very nice results were presented from coordinated observations
of several sources, including Mrk 421, between MAGIC and HESS
\cite{Mazin2}.
For Mrk 421, simultaneous data taken on the night of 18 Dec 2004
show a good agreement between the energy spectrum as measured by
MAGIC and HESS, where the spectral data can be fit by a single
power-law form across two orders of magnitude in energy.

\begin{figure}[ht]
\begin{center}
\includegraphics*[width=1.0\textwidth,angle=0,clip]{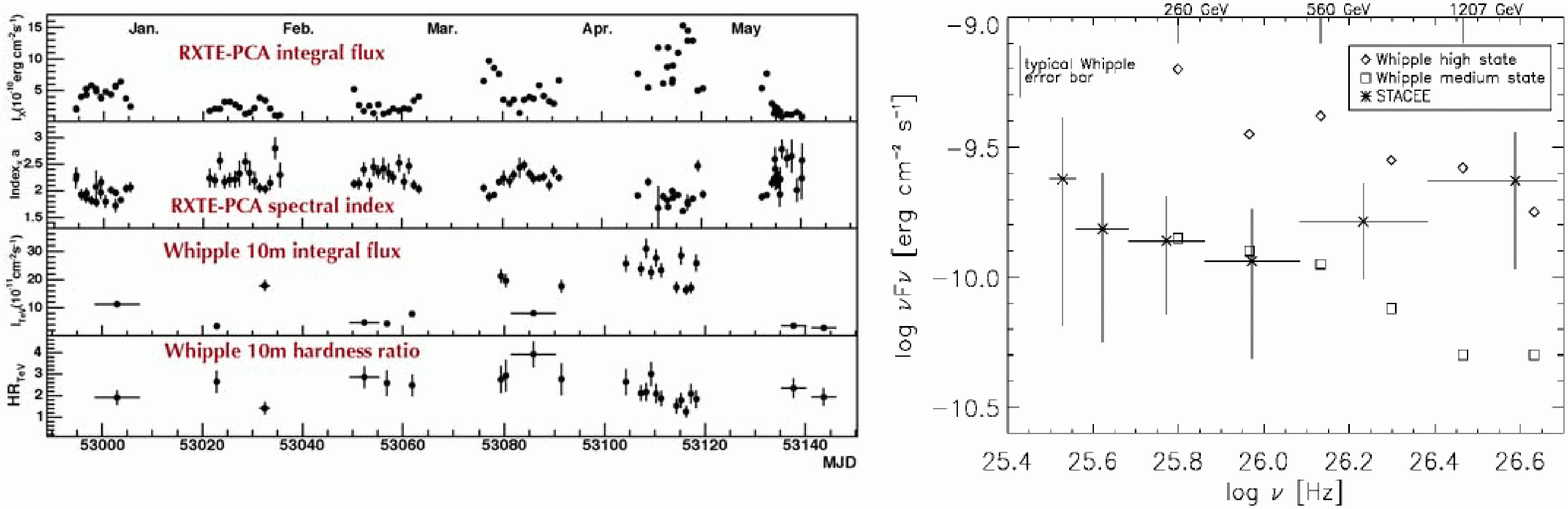}
\caption{\label {fig7}
Results from observations in 2004 of the blazar Markarian 421.
Left: light curves and spectral information for Mrk 421
for X-ray data from RXTE-PCA and VHE $\gamma$-ray data from
Whipple for the five month period between January and
May, 2005 \cite{Grube}.
Right: spectral energy distribution as measured by STACEE (stars)
at energies between 100 and 1500\,GeV for data in April 2004
\cite{Carson}.
Also shown are data from Whipple in medium and high flux emission
states.}
\end{center}
\end{figure}

Like Mrk 421, the source Markarian 501 (Mrk 501) is a nearby BL Lac object
that has exhibited a high degree of variability.
MAGIC reported on data taken in 2005 on Mrk 501 where the source was
in a flaring state \cite{Mirzoyan}.
On the one night of 1 July 2005, MAGIC detected the source with
a statistical significance of 41$\sigma$.
MAGIC was able to follow the flux variability on 2\,min time scales
and detected rapid variation on these time scales, as shown in
Figure~8.
These data may likely represent the shortest time variability ever
detected from any VHE $\gamma$-ray source and they place strong
constraints on the size of the $\gamma$-ray emission region.
MAGIC also reported on results from observations of the BL Lac
1ES 1959+650 \cite{Tonello}.
This source is interesting in that it is difficult
to detect in its low state and it has exhibited flaring at
VHE $\gamma$-ray energies that is unaccompanied by X-ray flaring
\cite{Orphan}.
The MAGIC data, taken in September and October 2004,
show a significant ($\sim 8\sigma$) detection, but no significant
time variability in the $\gamma$-ray flux.
The MAGIC observations
appear to sample the emission from the source in a low or quiescent
state.

\begin{figure}[ht]
\begin{center}
\includegraphics*[width=0.6\textwidth,angle=0,clip]{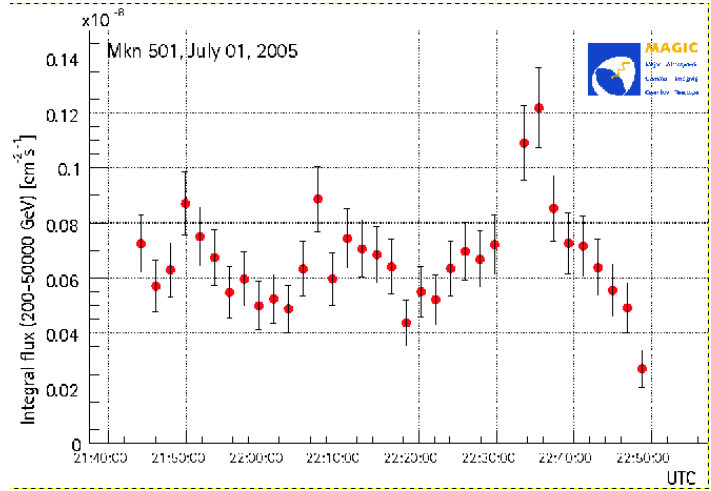}
\caption{\label {fig8}
Light curve for the blazar Markarian 501 for 01 July 2005 as
detected by MAGIC \cite{Mirzoyan}.
The $\gamma$-ray flux above 200\,GeV is plotted
in two minute time bins and
rapid variation is observed on that time scale.}
\end{center}
\end{figure}

PKS 2155-304 was the first extragalactic source detected in
the southern hemisphere.
Observations of this X-ray selected
BL Lac were reported by HESS \cite{Fontaine,Fontaine_update}
and CANGAROO \cite{Sakamoto}.
The HESS data, taken in 2003 and 2004, show a variety of interesting
features.  Although the source was detected overall at a very
high significance level ($\sim 100\sigma$), it was also detected
at low flux levels in every monthly period it was observed.
Multi-wavelength observations in 2003 show no correlation between
the X-ray flux as measured by RXTE-PCA and the HESS VHE $\gamma$-ray
flux; however, the data taken in 2004 do show a positive correlation.
The broad-band SED can be fit reasonably by both hadronic and
leptonic blazar models, depending on the choice of parameters and
on assumptions on the nature of the EBL.

Numerous experiments reported results on searches for other
extragalactic sources at VHE $\gamma$-ray energies.
AGN surveys have been initiated by both HESS \cite{Benbow,Benbow_update} and
MAGIC \cite{Meyer}, with the most important results presented
earlier.
STACEE reported on extensive observations of two low-frequency
peaked BL Lac (LBL) sources, W Comae and 3C 66A 
\cite{Mukherjee,Mukherjee_update},
the latter as part of a multi-wavelength campaign in 2003-2004.
W Comae is a source whose broadband SED is well understood so that
a detection at very high energies could potentially distinguish
between lepton and proton blazar models. 
The source 3C 66A is significantly more distant 
(nominal redshift of $z = 0.44$, but there is uncertainty in
this value) than any other AGN
detected to date by VHE telescopes.
STACEE did not measure a statistically significant $\gamma$-ray flux from
either source and reported integral upper limits in the 100-200\,GeV energy
range.
In addition to the detection of Mrk 421, PACT reported no evidence of flaring
from Mrk 501, H 1426+428, and ON 231 \cite{Bose}.
Similarly, TACTIC reported no detection of the BL Lacs 1ES 2344+514
\cite{Godambe} and H 1426+428 \cite{Thoudam}.
CANGAROO presented upper limits from observations of the radio galaxy
Centaurus-A that might be a mis-aligned BL Lac object \cite{Kabuki}.
Finally, the Perseus galaxy cluster was observed by Whipple for
$\sim 14\,$hrs in 2004-2005 \cite{Perkins}.
Galaxy clusters are considered to be promising
VHE targets where the $\gamma$-rays could originate from the
interactions of cosmic ray electrons or protons with material in the
cluster.
Whipple did not detect a significant $\gamma$-ray signal from Perseus
and placed flux upper limits from a number of the most luminous
radio sources within the cluster.

\section{OG 2.4: Gamma-Ray Bursts}

Gamma Ray Bursts (GRBs), discovered serendipitously in the
1960's, are among the most fascinating and enigmatic 
objects in all of high-energy astrophysics.
By demonstrating that GRBs were isotropic in their
arrival distribution,
BATSE, onboard the Compton Gamma Ray Observatory, gave credence
to the idea that GRBs were cosmological in origin.
However, the wide non-uniformity of GRBs, in terms of their
fluxes, light curves, and spectra, have made their characterization
difficult.
A key breakthrough was achieved later by the Beppo-SAX satellite
which produced accurate GRB positions and carried out X-ray follow-up
observations that enabled multi-wavelength observations of 
the afterglow emission from the class of longer (and softer) GRBs.
Redshift measurements of the optical counterparts of GRBs
indicate that the longer ones are indeed cosmological with a
typical redshift value of $z \sim 1 $.
Indeed, it is now realized these GRBs are likely the most energetic
explosions in the universe, with inferred outputs between
$10^{51}$ to $10^{54}$ ergs.
It is now generally accepted that GRBs are
a strongly beamed phenomena, and that when beaming is taken into
account their outputs generally cluster near $10^{51}$ergs.

Although a great deal is uncertain about the mechanisms behind
GRBs, a general paradigm regarding the long bursts has emerged
that starts with the explosion of a massive star (collapsar),
followed by the rapid formation of a highly relativistic jet.
Particle acceleration is carried out by relativistic shocks
in the jet, whereby internal shocks
produce the prompt emission and then, as the jet collides with
the surrounding medium, external shocks produce the afterglow
radiation.  Many questions regarding GRBs still remain
unanswered, including whether they produce significant amounts
of very high-energy $\gamma$-ray emission.
Numerous models predict strong emission at energies up to hundreds of
GeV and beyond, however, so far there has only been one
detection by EGRET of a photon above 10 GeV.
At this meeting, there were just a few papers discussing
GRB phenomenology \cite{Manchanda} or models \cite{Ellison2}.

Searches for VHE emission from GRBs or their afterglows
have been made by numerous ground-based instruments.
To date, there has been no convincing evidence for such emission.
The main satellite instruments that currently detect GRBs 
are HETE-2, INTEGRAL, and Swift.
Launched in November 2004, the Swift GRB mission is detecting
an unprecedented number of bursts with excellent positional information.
Ground-based instruments searching for VHE emission include
atmospheric Cherenkov telescopes, 
air shower arrays, and 
neutrino telescopes.

Air shower arrays have a high duty cycle and wide FOV that
enable them to observe GRBs during their phase of prompt emission.
However, the current instruments have substantial collection areas above
1\,TeV, reducing the likelihood that they would detect signals from
most of the long GRBs that are at cosmological distances.
Results were presented by Milagro based on data taken since 2000
\cite{Parkinson}. 
Data from 45 satellite-triggered bursts were analyzed; no
evidence for emission from any burst was found and limits on
the $\gamma$-ray fluence in the energy range between 0.25, and
25 TeV were obtained (assuming no attenuation of the signal
due to EBL absorption).
Milagro also reported on a generalized search for short duration
bursts, independent of a GRB trigger.
Data taken over a 2.3\,yr period for all positions in the sky
were analyzed for bursts
with durations ranging from 250\,$\mu$s to 40\,s;
the resulting burst limits were used along with simulations
to constrain the VHE spectrum of GRBs \cite{Noyes}.
Other results from air shower arrays include
first limits from the AGRO-YBJ instrument 
on a few GRBs from data taken in 2005 \cite{Girolamo} 
and from a general sky survey for steady sources and
transients \cite{Vernetto}
and limits at ultrahigh energies from data taken between 1996-2001
by the Andyrchy EAS array \cite{Smirnov}.
Reports were made on the feasibility of water Cherenkov
tanks for GRB detection as part of the Auger Project
\cite{Allard} and at a high altitude site in Mexico
\cite{Alvarez}.

Atmospheric Cherenkov telescopes have relatively low energy
thresholds and high sensitivity, but small fields of view.
They are thus well suited for follow-up observations of
the afterglows of triggered GRBs.
The MAGIC telescope has been designed to provide a very
rapid response to GRB alerts.
MAGIC reported on observations of a number of recent GRBs,
including one burst (GRB050713A) in which MAGIC started 
data-taking
only 40\,s after the burst event \cite{Bastieri,Bastieri_update}.
No evidence for VHE $\gamma$-ray emission was reported, but
this event indicates the significant potential of this
new generation of Cherenkov telescopes.
The STACEE solar array telescope reported on
observations of 14 GRBs
made over a several year period, including two Swift-triggered
bursts (GRB050402 and GRB050607)
in which data-taking started less than four minutes
after the GRB trigger \cite{Jarvis}.
Integral flux limits on GRB050607 were obtained for $\gamma$-ray
energies above 100\,GeV.

\section{OG 2.7: New Experiments and Instrumentation}

There were many contributions to ICRC 2005 in the area of
future experiments and instrumentation.
However, as discussed in the Introduction, this paper will only
briefly summarize these contributions because:
1) I am concentrating here on the present scientific status
of the field, and 2) there have been numerous meetings
discussing instrumentation and plans for future telescopes
(e.g. see \cite{Palaiseau}).

New experiments are under construction, or in the planning stages, 
in three areas: satellite-borne 
instruments, ground-based atmospheric Cherenkov telescopes, and ground-based
air shower arrays.

\subsection{New Satellite Instruments}
ASTROSAT is a UV/X-ray mission that is currently under development
by a number of institutions in India \cite{Agrawal}.
Scheduled to launch in 2007, the mission comprises five instruments,
spanning the UV, soft X-ray, and hard X-ray bands.
The UV Imaging Telescope (UVIT) has waveband coverage between 
130-320\,nm and an angular resolution of 1.8 arc-sec.
The Soft X-ray Telescope (SXT) images X-rays between 0.3-8.0\,keV over
field of view of 0.35$^\circ$.
The Large Area X-ray Proportional Counter (LAXPC) is a non-imaging
device covering a wide range of energies from 3-100\,keV with a
geometrical area of 10800\,cm$^2$.
The LAXPC will carry out variability and timing studies of X-ray sources
with modest energy resolution.
The CdZnTe Imager covers X-ray energies between 10-100\,keV
over a wide field of view, designed to carry out medium resolution
studies in the hard X-ray band.
The scanning sky monitor (SSM) is a position-sensitive proportional
counter to provide X-ray imaging between 2-10\,keV over a very wide
field of view. The science goals of ASTROSAT are varied, ranging from 
timing and spectral studies of AGN, SNRs, pulsars, binaries, and galaxy
clusters to detection of new X-ray transients.

GLAST is a future, major high-energy $\gamma$-ray mission that
is being developed by an international consortium 
and currently scheduled for launch in Fall 2007.
The mission consists of two instruments, the Large Area Telescope (LAT)
and the GLAST Burst Monitor (GBM).
The LAT comprises a large-area silicon strip tracker, a segmented
CsI calorimeter, and an anti-coincidence shield \cite{Moiseev}.
The LAT will
have greatly superior performance relative to EGRET
(i.e. an order of magnitude improvement in sensitivity).
Of particular importance for VHE astronomy, the LAT will have a
relatively constant effective area up to energies of 300\,GeV,
implying that many sources detected by GLAST will have measured
spectra that overlap with ground-based instruments.
The GBM is a gamma-ray burst detector that utilizes NaI and BGO
crystals to detect and localize bursts in a similar manner to BATSE
on the CGRO.  The GBM performance will be similar to BATSE but will
cover a wider energy range with a smaller collection area.
The construction of both the LAT and GBM instruments are largely complete
at the present time and full integration of the payload is starting.
The major science goals of GLAST are to discover new sources of high-energy
$\gamma$-rays (e.g. AGN, pulsars, SNRs, GRBs, and new objects), 
to more accurately measure spectra and positions of sources to understand
the mechanisms of high-energy particle acceleration, and to probe
dark matter and the early universe via observations of possible neutralino
signatures and of distant AGN and GRBs.

In addition to GLAST, the AGILE mission aims to study gamma-ray and X-ray
sources in the 30\,MeV - 30\,GeV and 10-40\,keV energy ranges, 
respectively.
AGILE is being developed by a group of Italian institutions for a launch
in the 2006-2007 time frame.  There was no contribution at ICRC 2005, but
details on the mission can be found at their website \cite{AGILE}.
There was also a contribution on CASTER, a coded-aperture imaging X-ray
telescope in the 10-600\,keV band that could be considered for 
NASA's Black Hole Finder Probe \cite{Cherry}.

\subsection{New Atmospheric Cherenkov Telescopes}
The VERITAS 
array will consist of four 12\,m diameter imaging
atmospheric Cherenkov telescopes,
to be deployed at a mountain site in southern Arizona, USA \cite{Holder}.
Each telescope images the Cherenkov light onto a 499-PMT array; the PMTs
are read out by a 500 MS/s Flash-ADC (FADC) system.
As shown in Figure~9, two telescopes have been constructed and deployed
at the Base Camp of the Whipple Observatory on Mt. Hopkins, Arizona.
The performance of the telescopes (mechanical, optical, electronic) all
meet or exceed the design specifications, and the first telescope has
been used during 2005 to make astrophysical detections \cite{Cogan}.
Figure~9 shows Cherenov events that have been reconstructed by two 
telescopes.
VERITAS is currently scheduled for full operation in late 2006.
When operational, VERITAS will be an important complement to MAGIC
in the northern hemisphere and to HESS and CANGAROO in the southern.

HAGAR is an atmospheric Cherenkov telescope array currently under
development for deployment at the high-altitude Hanle, India site
at 4200\,m elevation \cite{Chitnis}.
HAGAR will use the wavefront-sampling technique, employing seven
telescopes on a 50\,m hexagonal grid.  Each telescope will consist
of seven para-axially mounted 0.9\,m diameter mirrors with a single
PMT at the focus of each mirror.
Simulations indicate that HAGAR will have a low energy threshold
(below 100\,GeV) and moderate source sensitivity.
The first telescope has been constructed and deployed at Hanle
where tests are currently underway, with plans for a full
deployment by early 2007.

The MACE instrument is being designed as two very large (21\,m diameter)
imaging Cherenkov telescopes for future deployment at Hanle
\cite{Koul}.
The telescopes will be made up of 356 panels of mirrors, with each panel
consisting of either four or nine spherical mirror facets made from
diamond-machined aluminum alloy.
The MACE cameras will comprise 832 pixels, covering a field of view
of 4$^\circ$ x 4$^\circ$.
With its large size and high-altitude site, MACE should have a trigger
energy threshold well below 50\,GeV.
Designs for MACE are currently in the advanced stages with a 4-5\,yr
construction period envisioned.

The first phase of HESS, consisting of four 12\,m imaging atmospheric
Cherenkov telescopes was comissioned in late 2003.
Planning is underway to construct HESS Phase II \cite{Vincent};
the major component of the upgrade will be a very large 28\,m diamter
telescope that will be deployed at the center of the
HESS-I array in Namibia.
The new telescope will significantly improve the energy threshold
of HESS and increase its sensitivity.
Various options for the reflector are being considered;
initial plans call for a rectangular dish (32\,m x 24\,m) with 850
mirror facets and a parabolic shape.
The very large camera will consist of 2048 pixels with a front-end
readout that will be redesigned from HESS-I to accomodate the
higher trigger rates ($\sim 3\,$kHZ).
Construction on HESS-II has started with a planned completion time
of approximately three years.

MAGIC is also in the process of an upgrade to the original instrument.
The major component of MAGIC Phase II will be a second 17\,m
diameter reflector \cite{Teshima}.
Other improvements will be the replacement of the replacement of the
standard camera with hybrid photodetectors (HPDs) that have a
quantum efficiency near 50\% and new FADCs with a sample rate of
2.5\,GS/s.
These various improvements should lower the energy threshold of MAGIC
and improve its sensitivity, especially at energies below 100\,GeV.
Construction of the second telescope is now well underway with a planned
completion date of sometime in 2007.

\subsection{Air Shower Arrays}
As shown in Table~1, the ARGO-YBJ experiment is an air shower array
deployed at the high altitude site of Yangbajing, Tibet.
The experiment will eventually consist of a large carpet
of resistive plate chambers (RPCs) covering a total area
of 6400\,m$^2$ \cite{Girolamo}.
The RPCs are divided into 18480 pads of size 0.56\,cm x 0.62\,cm that
provide the space-time pattern for measuring the wavefront of the
air shower.
The experiment is still under construction, but a large portion ($> 50\,$\%)
is already installed and operational.
ARGO-YBJ has presented early physics results on a variety of topics
at this meeting, including a sky-survey and a GRB search.
Plans call for the full experiment to be operational in 2006.

A future air shower array using the water Cherenkov technique
was also discussed \cite{Sinnis2}.
The HAWC experiment would consist of a 300\,m x 300\,m pond
of water that would be instrumented by a large number of PMTs.
A key attribute of HAWC would be its high altitude location
($> 4000$\,m elevation) to permit a substantial
lowering of the energy threshold relative to Milagro.
Detector and sensitivity studies are underway to characterize
HAWC.
The design of an earlier version of the experiment, called mini-HAWC, 
that could utilize the existing 900 PMTs from Milagro, is also being actively 
pursued.

\begin{figure}[ht]
\begin{center}
\includegraphics*[width=0.95\textwidth,angle=0,clip]{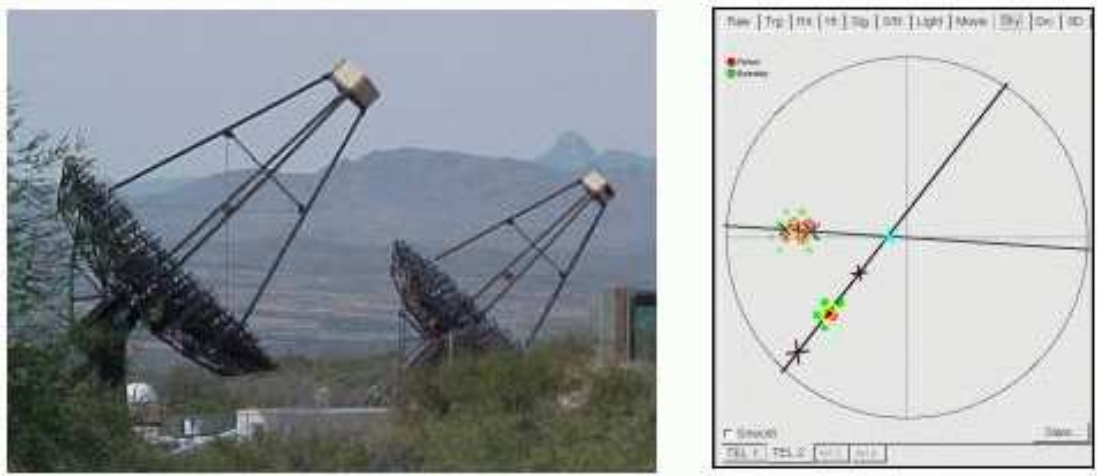}
\caption{\label {fig9}
The VERITAS atmospheric Cherenkov telescope array.
Left: the first two telescopes VERITAS, operating 
at the Base Camp of Mt. Hopkins, Arizona, USA \cite{Holder}.
VERITAS will consist of four 12\,m diameter telescopes, each
equipped with a 499 pixel camera.
Right: A stereo Cherenkov event taken with the first
two telescopes.}
\end{center}
\end{figure}

\pagebreak

\section{Conclusions}

The last two years have been exciting times for very high energy (VHE)
$\gamma$-ray
astronomy.
Some of the most significant developments are the following:

\begin{enumerate} 

\item A new generation of atmospheric Cherenkov telescopes 
has yielded outstanding
results, including the discovery of many more sources of VHE radiation.

\item The spatial and spectral properties of VHE
sources are being measured with a precision that is unique in 
gamma-ray
astronomy and that, in many ways, exceeds what was done with the Compton
Gamma-Ray Observatory, a Great Observatory of NASA.

\item We have learned that the Galactic plane is rich in the number and type of
VHE sources, with supernova remnants and pulsar wind nebulae
now firmly established as important
VHE emitters.  Other new sources include a binary pulsar, a microquasar,
and the mysterious Galactic center.

\item We have identified a number of new sources that do not have obvious
counterparts to objects at other wavelengths. One of these is a broad source
in the Cygnus region of the Galactic plane.
We are starting the investigation of a new class (or classes) of
astrophysical objects that are bright at TeV energies, but dim at 
other wavebands.

\item The discovery of four new 
blazars increases the extragalactic source count
and pushes our discovery space out to greater redshift values.
The fact that the most distant VHE sources have unbroken power-law spectra to
the highest energies detected may signal that the universe is more transparent
to VHE photons than previously suspected.

\item New experiments on the ground and in space should continue the rapid
and exciting development of VHE astrophysics.

\end{enumerate}

\section{Acknowledgments}
The author thanks the organizers of ICRC 2005 for the invitation to give
this rapporteur talk and acknowledges
the assistance and encouragement
of Sunil Gupta and Suresh Tonwar.
The talk would not have been possible without considerable help from
the major experimental groups in ground-based gamma-ray astronomy worldwide,
most notably the AGRO-YBJ, CANGAROO, HESS, MAGIC, Milagro, STACEE, TACTIC,
Tibet, and VERITAS collaborations.
Special thanks is due to Rudolf Bock, Werner Hofmann, Masaki Mori, and
Gus Sinnis who went to great lengths before (and during) the meeting
to provide copies of the presentations
from their respective groups.


\begin{thebibliography}{99}
\bibitem{Grading}
Source detection are typically classified as being
solid (A), probable (B), and possible (C).
A solid source is a $> 5\sigma$
detection by two or more instruments,
a probable source is a $> 5\sigma$
detection by one instrument, and a possible
source is a reported detection by one instrument.
This paper discusses solid and probable detections.
\end{thebibliography}
\end{document}